\documentclass[%
preprint,
superscriptaddress,
nofootinbib,
amsmath,amssymb,
aps,
longbibliography,
prapplied,
]{revtex4-2}

\usepackage{graphicx}
\usepackage{amssymb,amsmath,amsfonts, latexsym}
\usepackage{multirow}
\usepackage{xcolor}
\usepackage[top=2.5cm, bottom=2.5cm, left=2cm, right=2cm]{geometry}
\usepackage{threeparttable}

\begin{document}

\newcommand{\pjump}{[\![P]\!]}

\title{Energy conversion from heat to electricity by highly reversible phase-transforming ferroelectrics}
\author{Chenbo Zhang}
\affiliation{Department of Mechanical and Aerospace Engineering, Hong Kong University of Science and Technology, Clear Water Bay, Hong Kong}
\affiliation{HKUST Jockey Club Institute for Advanced Study, Hong Kong University of Science and Technology, Clear Water Bay, Hong Kong}
\author{Zhuohui Zeng}
\affiliation{Department of Mechanical and Aerospace Engineering, Hong Kong University of Science and Technology, Clear Water Bay, Hong Kong}
\author{Zeyuan Zhu}
\affiliation{Department of Mechanical and Aerospace Engineering, Hong Kong University of Science and Technology, Clear Water Bay, Hong Kong}
\author{Nobumichi Tamura}
\affiliation{Advanced Light Source, Lawrence Berkeley National Laboratory, CA 94720, USA}
\author{Xian Chen}
\affiliation{Department of Mechanical and Aerospace Engineering, Hong Kong University of Science and Technology, Clear Water Bay, Hong Kong}
\email{xianchen@ust.hk}

\baselineskip20pt

\begin{abstract}
    Searching for performant multiferroic materials attracts general research interests in energy science as they have been increasingly exploited as the conversion media among thermal, electric, magnetic and mechanical energies by using their temperature-dependent ferroic properties. Here we report a material development strategy that guides us to discover a reversible phase-transforming ferroelectric material exhibiting enduring energy harvesting from small temperature differences. The material satisfies the crystallographic compatibility condition between polar and nonpolar phases, which shows only 2.5$^\circ$C thermal hysteresis and high figure of merit. It stably generates 15$\mu$A electricity in consecutive thermodynamic cycles in absence of any bias fields. We demonstrate our device to consistently generate 6$\mu$A/cm$^2$ current density near 100$^\circ$C over 540 complete phase transformation cycles without any electric and functional degradation. 
    The energy conversion device can light up a LED directly without attaching an external power source. This promising material candidate brings the low-grade waste heat harvesting closer to a practical realization, \emph{e.g.} small temperature fluctuations around the water boiling point can be considered as a clean energy source. 
\end{abstract}


\maketitle


\section{Introduction}
Waste heat is inevitably generated from a plethora of processes, such as burning fuel, running computers, and solar radiation. The thermal energy that reaches temperatures of around 100$^\circ$C is abundant and has drastically increased in recent years from urban and industrial activities. As such, identifying methods to recycle and harvest waste heat is among the most pressing challenges in energy science. While thermal to electricity energy conversion technologies have been advanced by a large margin in thermoelectric devices in recent years, their applications to the energy harvesting from low-grade heat (\emph{e.g.} $< 150^\circ$C) are highly hindered by the low ZT value in the low temperature regime.   

Electrically polarized crystals naturally exhibit the pyroelectric effect: a change in temperature causes a change in its spontaneous electrical polarization.
The earliest documented observation of pyroelectric effect in stones such as tourmaline can date back to ancient Greece. \cite{lang1974} In fact, the pyroelectricity exists in all non-centrosymmetric crystal structures. Among 7 crystal systems, only cubic system is centrosymmetric, which implies that any non-cubic structures may exhibit pyroelectricity below their Curie temperatures. 
Since the 1960s, pyroelectric materials have attracted profound attention due to their suitability for thermal sensing and thermal imaging \cite{whatmore1986, bauer1989simple}. Later, some theoretical models were proposed by Olsen, Evans and Drummond to explore the heat-to-electricity conversion as utilizing these materials as the dielectric layer in capacitors, called dielectric power converter (DPC) \cite{drummond1979, olsen1983, olsen1985cycle}. 
Their work underlies the thermodynamics of energy conversion by pyroelectric effect. 
Since then, this idea has been popularized as an important energy harvesting approach for recycling the low grade waste heat, which was demonstrated in relaxor ferroelectrics \cite{olsen1983, sebald2007, sebald2009} by the electric Ericsson cycle \cite{olsen1985cycle}. The pyroelectric current is produced in the isoelectric processes when the system is heated up and cooled down under an alternative applied voltages. In Olsen's early demonstration \cite{olsen1985cycle}, the pyroelectric converter was operated at 200V and 700V in a temperature range from 140$^\circ$C to 160$^\circ$C. In Olsen's design, a drawback is the risk of applying high DC voltages during energy conversion cycles. 
As the nano-fabrication technologies flourished, the Olsen cycles have been realized in thin film devices. \cite{navid2011, yang2012, bhatia2014high, pandya2018nm} The applied bias voltage between two isoelectric processes of the Ericsson cycle is reduced to 2V and 8V for thin film pyroelectric capacitors \cite{pandya2018nm, sharma2021lead}. A main problem of these latest devices is the confusion where the electricity comes from. When an external electric source is hooked up during energy conversion, the pyroelectric current and the current caused by switching the external voltage are both collected in the circuit across the same load resistor. In some design of pyroelectric converter, the electric energies generated from the thermal fluctuations and from the electric field alternation are both accounted as the harvest energy \cite{pandya2018nm}. 

In absent of applied electric bias field, the pyroelectric effect becomes very weak in common relaxor thin films, e.g. 0.055 $\mu$C/(cm$^2$ K) in PMN–0.32PT within the working temperature range for energy conversion \cite{pandya2018nm}. This is one of the obstacles preventing pyroelectric energy conversion from broad commercialization. Recent advances in innovative thermal management \cite{pandya2017prapplied, pandya2018nm} and exquisite nano-devices \cite{yang2012,ji2019piezo, sharma2021lead} have pushed the device performance towards its current limit, that is up to several hundreds of nanoamperes \cite{yang2012, bhatia2014high, pandya2018nm, ji2019piezo, sharma2021lead}. But the ambient thermal fluctuations in nature can never reach the heating and cooling frequencies modeled in some reported devices \cite{pandya2018nm,sharma2021lead}.
This field urgently calls for new materials to intrinsically improve the multiferroic-thermal response. 

\section{Material development strategies}
In relaxor ferroelectrics, pyroelectric effect is characterized by the pyroelectric coefficient defined as $dP/dT$, where $P$ is the spontaneous polarization and $T$ is the temperature. 
Conventionally, searching for a large pyroelectric coefficient underlies the material design strategy. Therefore, it is common to utilize the lead-based ferroelectrics such as PZT and PMN-PT \cite{sebald2007, sebald2009, pandya2018nm}. When a sharp phase transformation between polar and non-polar structures occurs, the sudden drop in polarization within quite a narrow temperature interval gives rise to a singular $dP/dT$ value. The phase transformations of barium titanate based ferroelectrics belong to this case, whose transition temperature is around 120$^\circ$C. Compared to the relaxor ferroelectrics, the first-order phase-transforming barium titanate based ferroelectrics are more suitable for the conversion from heat to electricity, especially in the low-grade temperature regime. Since there is no spontaneous polarization in the cubic phase, we will no longer use the terminology of pyroelectric energy conversion as for the heat-to-electricity conversion by first-order phase transformation in this paper. We will continue using pyroelectric coefficient defined as $dP/dT$ to express the temperature-dependent polarization variation, but $dP/dT$ is a function of temperature, which becomes singular at the phase transition temperature.
As an example, the single crystal BaTiO$_3$ undergoes a first-order phase transformation from tetragonal to cubic, whose $dP/dT$ at transition temperature is calculated as $\frac{\Delta P_\textup{r}}{\Delta T} \approx 1\mu$C/cm$^{2}$K from experimental measurements \cite{moya2013}. That is about twice of the highest $dP/dT$ reported in lead-based relaxor PMN-0.32PT single crystal along [111] polar axis at its transition temperature (180 $^\circ$C) \cite{kumar2004}. Note that all reported working temperatures of relaxor pyroelectric energy conversion is below 160$^\circ$C (\emph{i.e.} within a single phase regime). It means that relaxor devices are operated under much lower performance for energy conversion. \cite{yang2012, bhatia2014high, pandya2018nm} 

Utilizing the first-order phase transformation, people has indeed demonstrated 
promising energy conversion performance using single crystal BaTiO$_3$ by a $\pm 5^\circ$C temperature fluctuations near the transition temperature. \cite{bucsek2019} 
Reserved energies are more abundant in small thermal fluctuations than in a static temperature gradient. These rich thermal energies mainly oscillate at $\sim 2^\circ$C/s (0.5Hz) rate around $100^\circ$C from boiling water, running computers and industrial waste heat. \cite{zhang2020thermal} 
Hence, finding the ferroelectric material undergoing first-order phase transformation at a proper temperature is a core strategy of material optimization. Compared to relaxor ferroelectrics, the barium titanate and its sibling derivatives are the better candidates for energy conversion. 

From energetic analysis, the cost of the energy conversion is the heat in the heating half cycle or the work required to cool the active material in the cooling half cycle, while the return is the change of polarization stored as the electrostatic energy. 
The ratio of pyroelectric coefficient to the heat capacity at a fixed bias field is a natural choice of figure of merit, which is in fact proposed for infra-red sensors, such as $F_i = (1/c_E)dP/dT$, $F_v = (1/\epsilon_{33} c_E)dP/dT$ where $c_E$ is the volume specific heat and $\epsilon_{33}$ is the permittivity along the polarized direction. \cite{lang2001book, bowen2014} Both $F_i$ (current) and $F_v$ (voltage) represent the electric responsivity of a sensing device to the ambient temperature variations, not considering a thermodynamic cycle of energy harvesting. Another figure of merit for pyroelectric energy harvesting is derived from the analogy of the piezoelectric energy conversion, defined as $F_E = (1/\epsilon_{33})(dP/dT)^2$ \cite{sebald2006fom}. Compared to the figures of merit $F_v$ and $F_i$, the harvesting figure of merit, $F_E$, does not consider the heat capacity. Although it was widely used for device designs \cite{sebald2008pyroelectric, mangalam2013improved, pandya2018nm}, the definition of $F_E$ does not take into account the transient nature of heat transfer and the singularity of the transport properties in the vicinity of phase transition temperature. 

Across a phase transformation, the effective heat capacity integrated over temperature is the latent heat $\ell$, while the integrated $dP/dT$ is the jump of polarization $\pjump$.
Based on thermodynamic analysis of a energy harvesting cycle without application of a bias field \cite{zhang2019power}, we propose a figure of merit considering the conversion in a temperature range covering a complete loop of phase transformation. That is,
\begin{equation}\label{eq:fom}
    \textup{FOM} = \dfrac{\kappa\pjump}{\ell},
\end{equation}
where $\kappa$ is the maximum value of $\vert dP/dT\vert$ near the transformation temperature, $\pjump$ is the jump of polarization across the phase transformation corresponding to the latent heat $\ell$. Eq. \eqref{eq:fom} rationalizes the searching of ferroelectric materials specially desirable for conversion from small temperature fluctuations to electricity by phase transformation. It is one of the key factors considered in this paper for material development. 

To realize the cyclic transformations in a narrow temperature interval for energy conversion, the thermal hysteresis and functional degradation are unprecedented factors for the material design. It has been theorized that the transformation hysteresis and reversibility strongly rely on the crystallographic compatibility, that is a tensor-metric condition on lattice parameters of both phases.\cite{ball1989fine, chen2013cc} Mathematically, the primary criterion of compatibility is $\lambda_2 = 1$ where $\lambda_2$ is the middle eigenvalue of the transformation stretch tensor.\cite{chen2013cc} The value of $\lambda_2$ can be tuned by the lattice parameters of initial and final phases. It has been successfully used as a search index for low-hysteresis shape memory alloys \cite{cui2006, zarnetta2010, song2013, chluba2015} and transforming oxides with enhanced structural reversibility \cite{pang2019reduced, jetter2019, liang2020, wegner2020}. Some recent discoveries have shown that ferroic properties such as electromagnetic property \cite{el2019PRM}, ferroelectric property \cite{wegner2020} and elasto/magneto caloric property \cite{zhao2017acta} are closely related to the crystallographic compatibility, and can be optimized when lattice parameters meet the $\lambda_2 = 1$ condition. Since ferroic properties are keenly sensitive to crystal structures, searching for proper dopants to finely tune the lattice parameters of barium titanate is another important design strategy.  

\section{Material development}

Considering the working temperature regime (about 100$^\circ$C), the non-lead toxicity and  the first-order phase transformation, we select three dopants Ca, Zr, Ce to develop barium titanate based ferroelectrics, named as tridoped BT. The reasons are following. The Ca addition at 5 at.\% is to ensure a reversible tetragonal (ferroelectric) to cubic (paraelectric) transformation below 120$^\circ$C, according to the phase diagrams \cite{liu2015BCZT, coondoo2021}. 
Preliminary studies \cite{zhang2019power, zhang2020leakage, wegner2020} have shown that the addition of Zr increases the polarization jump across the phase transformation, potentially boosting the FOM.
The charge leakage in ferroelectrics is unavoidable and non-negligible, especially at elevated temperature \cite{nagaraj1999leakage, pabst2007leakage, zhang2020leakage}. This highly hinders the long-lasting energy conversion performance due to the electric degradation, discussed in reference \cite{zhang2020leakage}. Experiments \cite{wang2005leak, zhang2020leakage} show that Ce is an effective dopant to prevent the charge leakage in barium titanate system. Compared to BaTiO$_3$ and Ba(Zr, Ti)O$_3$ systems, only 1 at.\% Ce doping can reduce the leakage current density by 1$\sim 3$ orders of magnitude. 
The subtle lattice parameters adjustments can be achieved by tuning compositions of the dopants Zr and Ce within the range of 0.1 at.\% $\sim 1$ at.\%. 

\subsection{Synthesis and characterizations}
A series of Ba$_{0.95}$Ca$_{0.05}$Ti$_{1-x-y}$Zr$_{x}$Ce$_{y}$O$_3$ oxides were synthesized by solid-state reaction and densification process to obtain homogeneous polycrystalline rods of high density.
The powders of CaCO$_3$(Alfa Aesar, 99.5\%), CeO$_2$(Alfa Aesar, 99.9\%), ZrO$_2$(Sigma Aldrich, 99\%), BaCO$_3$ (Alfa Aesar, 99.8\%) and TiO$_2$(Alfa Aesar, 99.8\%) were weighed and well mixed according to the stoichiometric formulation 
Ba$_{0.95}$Ca$_{0.05}$Ti$_{1-x-y}$Zr$_x$Ce$_y$O$_3$, 
for ($x$,$y$) = (0.001, 0.005), (0.005, 0.005), (0.01, 0.005), (0.01, 0.01), and (0.01, 0.001). 
Then, the powder mixture was dissolved in the ethanol solvent, which was milled by zirconia balls in a planetary ball miller at 630rpm for 24hrs. The ball-milled solution was dried and calcined at 1000$^\circ$C for 10 hours to obtain the tridoped barium titanate powder through solid state reaction. Under a 30MPa hydrostatic pressure at room temperature, the powder was hot-pressed for 30 minutes to form a rod-shape green body with size $10$mm $\times \phi 6$mm.
The sintering process is divided into two steps by four-mirror Infrared furnace (Quantum Design IRF11-001-00): 1) densification of feed and seed rods \cite{zhang2020leakage}; 2) grain-coarsening by floating-zone method \cite{balbashov1981, kimura1992}. The densification was conducted as a pre-sintering process by passing through the focal point of the four mirrors slowly with the axial speed 3mm/hr and the angular speed 3rpm. With much less sintering time, all pre-sintered rods show high densities ($5.4 \sim 5.6$g/cm$^3$) that is $> 90$\% of theoretical density of BaTiO$_3$.   The compositions of tridoped BT samples are characterized by the Energy-dispersive X-ray Spectroscopy analysis by JEOL 6390. The variation of the normalized intensity was characterized for elements Ba, Ti, O, Ca, Zr and Ce along a 1.2 mm line on surface for all samples. The atomic weight percentages were analyzed based by the spectra of the polycrystalline tridoped BT samples. The compositions are consistent with the nominal compositions, and the variations are sufficiently small with respective to their mean values. 

\begin{figure}
    \centering
    \includegraphics[width=0.37\textwidth]{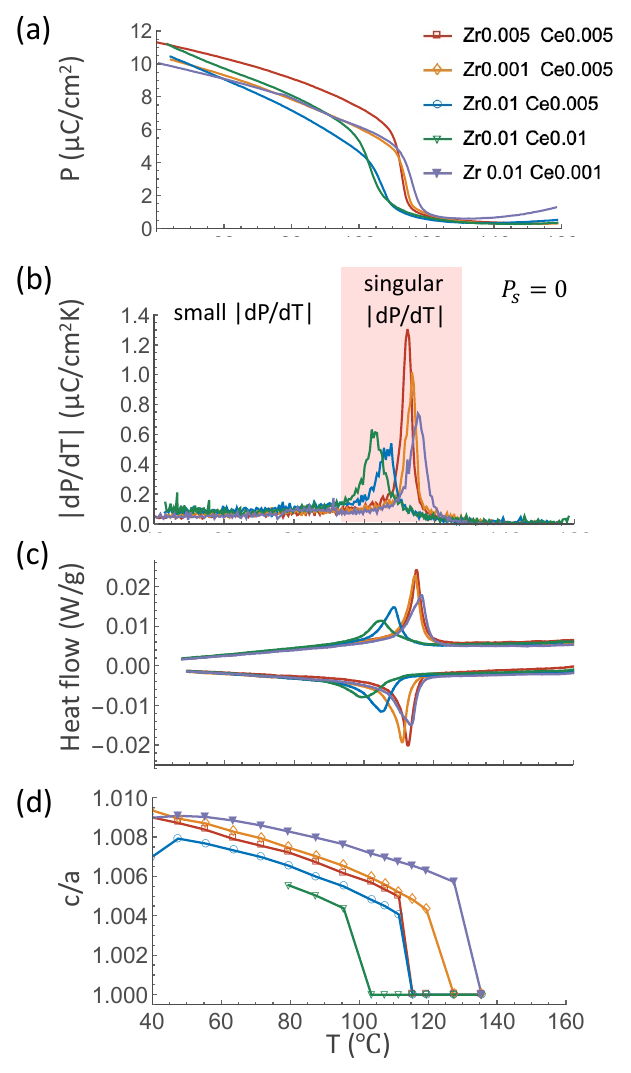}
    \caption{Ferroelectric, thermal and structural characterizations of first-order phase transformation in polycrystalline tridoped BT crystals. (a) and (b) Temperature dependent polarization and absolute value of pyroelectric coefficent $dP/dT$. (c) Thermal analysis of reversible phase transformations by DSC. (d) Temperature dependent $c/a$ ratio by neutron diffraction experiments. \label{fig:ferro}}
\end{figure}

The temperature dependent ferroic properties were characterized by the aix ACCT TF2000E ferroelectric analyzer at a temperature step 0.5$^\circ$C.
Figure \ref{fig:ferro} (a) and (b) show the temperature dependent spontaneous polarization and corresponding $|dP/dT|$ of five compositions, labeled as ``Zr$x$ Ce$y$'' for 
$(x, y) = (0.005, 0.005)$, $(0.001, 0.005)$, $(0.01, 0.005)$, $(0.01, 0.01)$ and $(0.01, 0.001)$.

All of the tridoped BT show large $\pjump$ and steep $dP/dT$ slope within 100 to 120$^\circ$C, among which the Zr0.005 Ce0.005 specimen gives the largest $\pjump = 7.58\mu$C/cm$^2$ and $\kappa = 1.31\mu$C/cm$^{2}$K.
The thermal analysis was conducted by differential scanning calorimetry (DSC) using TA Instruments DSC-250 in Figure \ref{fig:ferro}(c). The result reaffirms that they all undergo first-order phase transformations as big peaks of the heat flow are detected during heating and cooling through the transformation temperatures. 
Table \ref{tab:fom} summarizes the ferroelectric properties and FOM defined in Eq. \eqref{eq:fom} for these five samples compared with typical ferroelectric poly/single crystals reported in literature. Among them, Zr0.005 Ce0.005 gives the largest FOM, that is 1.52 $\mu$C$^2$/JcmK in polycrystal and 3.55 $\mu$C$^2$/JcmK in single crystal. 

\begin{table*}
    \caption{Summary of energy conversion properties and compatibility of samples developed in this work and important ferroelectrics reported in literature. The $\pjump$ is calculated for a 20$^\circ$C temperature interval near 100$^\circ$C.   \label{tab:fom}}
    \begin{tabular}{|l|c|c|c|c|c|}
         \hline
         \multirow{2}{*}{Materials} & $\pjump$ & $\kappa$ & \multirow{2}{*}{$\ell$ (J/cm$^3$)} & FOM & Temp. range\\
         & ($\mu$C/cm$^2$) & ($\mu$C/cm$^{2}$K) & & ($\mu$C$^2$/J cm K) & ($^\circ$C)\\ \hline
         Zr0.005 Ce0.005 & 7.58 & 1.31 & 6.55 & 1.52 & \multirow{2}{*}{105-125}\\
         Zr0.005 Ce0.005 sc & 11.68 & 1.93 & 6.34 & 3.55 & \\\hline
         Zr0.001 Ce0.005 & 6.11 & 1.02 & 6.20 & 1.00 & 105-125 \\ \hline
         Zr0.01 Ce0.005 & 5.58 & 0.54 & 5.33 & 0.57 & 100-120 \\ \hline
         Zr0.01 Ce0.01 & 6.48 & 0.63 & 5.67 & 0.72 & 100-120 \\ \hline
         Zr0.01 Ce0.001 & 5.81 & 0.74 & 6.53 & 0.66 & \multirow{2}{*}{105-125} \\
         Zr0.01 Ce0.001 coarse & 6.27 & 1.01 & 6.28 & 1.01 & \\\hline\hline
         BaTiO$_3$ sc \cite{moya2013} & 6.0 & 1.0 & 5.69 & 1.05 & 115-135 \\\hline
         Ba(Ti, Zr$_{0.017}$)O$_3$ poly\cite{wegner2020} & 2.5 & 0.58 & $\sim 5 \text{ to } 6$ & 0.24 to 0.29 & 120-140\\\hline
         PMN - 0.32PT sc & 1.56\cite{kumar2004}  & 0.08\cite{kumar2004}  & 7.5 \footnote{The transferred heat is calculated as $c_E \Delta T$ for specific heat $c_E = 0.3$J/cm$^3$K \cite{lau2008lead}.} & 0.014 &100-125\\\hline
         PMN - PT thin film &  & 0.055\cite{pandya2018nm} & & n.a.\footnote{There is no reported thermal analysis of this system.} & 100-120\\\hline
    \end{tabular}
\end{table*}

\subsection{Correlation between the transport properties and crystallographic compatibility}

\begin{figure*}
    \centering
    \includegraphics[width=0.8\textwidth]{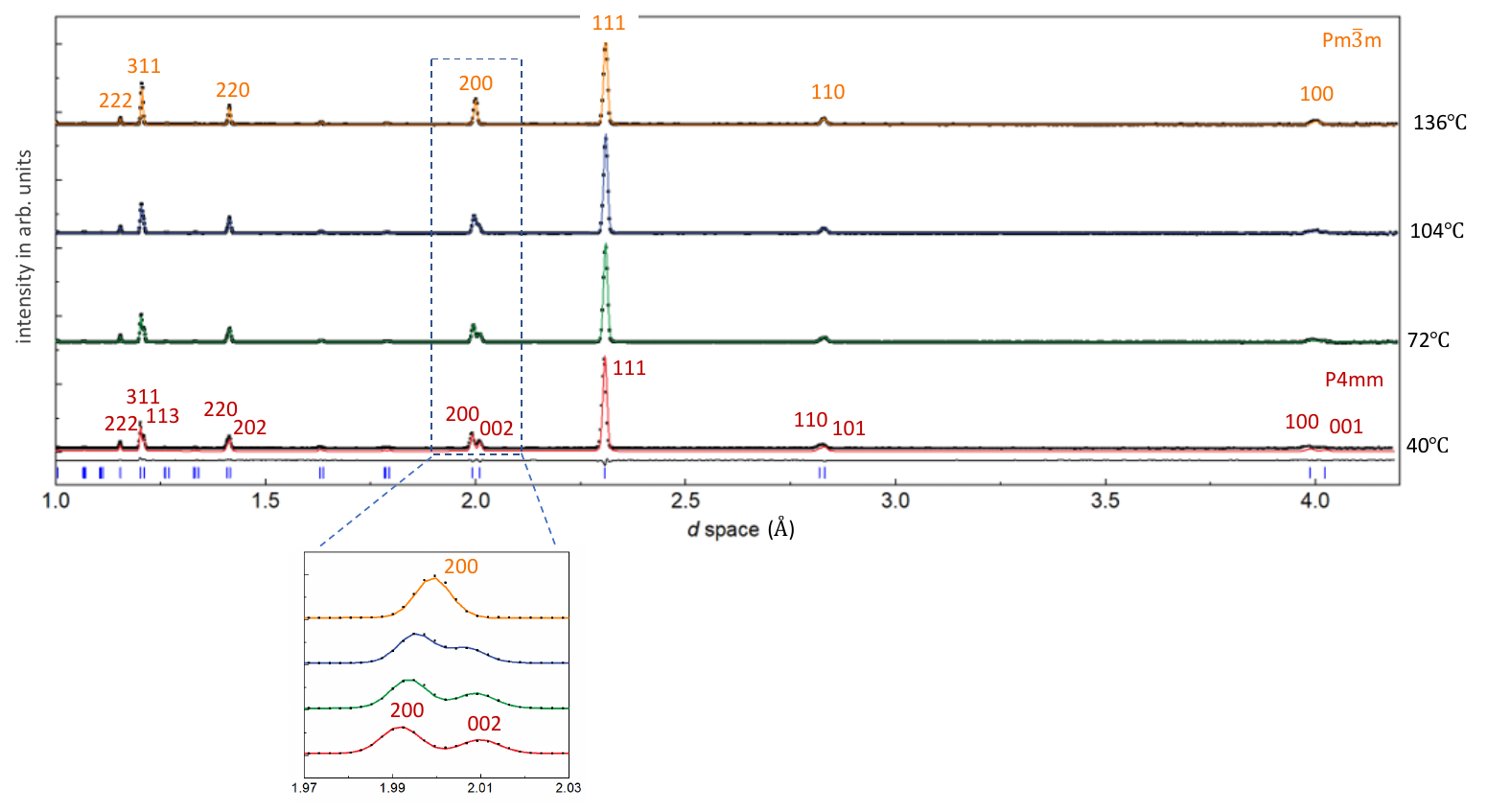}
    \caption{Neutron diffraction pattern of Zr0.005 Ce0.005 polycrystalline bulk sample at elevated temperatures from 40$^\circ$C to 140$^\circ$C indexed by $Pm\bar3m$ (cubic) and $P4mm$ (tetragonal) symmetries. }
    \label{fig:neutron}
\end{figure*}

The lattice parameters were measured by neutron diffraction at the general purpose powder diffractometer beamline (GPPD) at China Spallation Neutron Source, suitable for structural characterization of large bulk samples.\cite{chen2018general} We conducted the temperature varying diffraction experiments for densified polycrystalline rods ($\phi 5 \times 50$mm$^3$) from 40$^\circ$C to $140^\circ$C at a step of 10$^\circ$C per scan. Before each of the $\theta-2\theta$ scans, the bulk sample was set in an isothermal environment until the temperature reading was stable. The cubic (high temperature phase) and tetragonal (low temperature phase) lattice parameters were refined by $Pm\bar3 m$ and $P4mm$ space groups by the Rietveld refinement using GSAS suite with EXPGUI \cite{toby2001expgui}. Same procedure was conducted for rest four bulk samples in different compositions. The tetragonality defined as the $c/a$ ratio is presented in Figure \ref{fig:ferro}(d) where the lattice parameters were refined based on the diffraction patterns at different temperatures (e.g. Figure \ref{fig:neutron}). 

Stark discontinuity of tetragonality exists in all compositions (Figure \ref{fig:ferro}d), which evidences the first-order structural phase transformation in consistent with the abrupt jump of polarization (Figure \ref{fig:ferro}a), the singularity of pyroelectric coefficient (Figure \ref{fig:ferro}b) and the big heat flow peaks in DSC measurements (Figure \ref{fig:ferro}c). 

\begin{figure}
    \centering
    \includegraphics[width=0.33\textwidth]{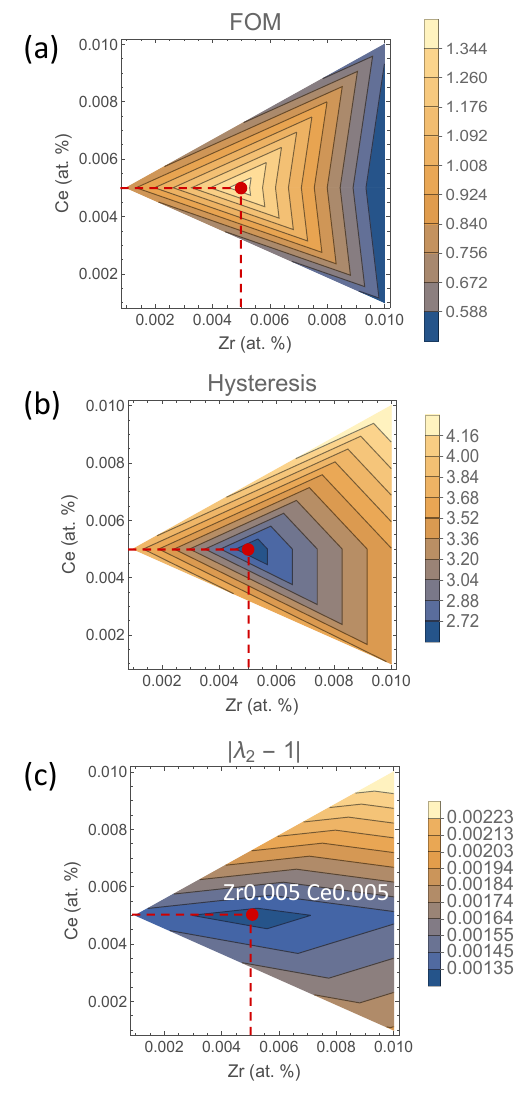}
    \caption{Contours of (a) FOM, (b) thermal hysteresis and (c) compatibility condition in two dimensional composition space interpreted from experimental data. }
    \label{fig:corr}
\end{figure}

The transformation stretch tensor between tetragonal and cubic phases is calculated as
\begin{equation}
    \mathbf U = \begin{bmatrix}\frac{a}{a_0} & 0 & 0\\0&\frac{a}{a_0}&0\\0&0&\frac{c}{a_0}\end{bmatrix}.
\end{equation}
where $a_0$ is the lattice parameter of cubic phase, $a$, $c$ are the lattice parameters of tetragonal ferroelectric phase. The primary compatibility indexer $\lambda_2$ is calculated as the middle eigenvalue of $\mathbf U$, directly computed as $\lambda_2 = \frac{a}{a_0}$. It has been theorized that the distance of $\lambda_2$ to 1 underlies the height of energy barrier of a symmetry-breaking transformation \cite{zhang2009energy}. It has been shown in many phase-transforming systems that the thermal hysteresis and reversibility of a solid phase transformation strongly relies on the compatibility condition.\cite{ball1989fine, cui2006, zarnetta2010, song2013}

\begin{table*}
 \caption{Lattice parameters of polycrystalline tridoped BT samples measured by neutron diffraction corresponding to their compatibility indexer $\lambda_2$. \label{tab:latt}}
    \begin{tabular}{c|c|c|c|c}
         Materials & $a_0 (\AA)$ after $T_c$ & $a (\AA)$ before $T_c$ & $c (\AA)$ before $T_c$ & $\lambda_2$ \\\hline
        Zr0.005 Ce0.005 & 3.99908 & 3.99388 & 4.01373 & 0.9987 \\
        Zr0.001 Ce0.005 & 4.00345 & 3.99744 & 4.01506 & 0.9985 \\
        Zr0.01 Ce0.005 & 4.00716 & 4.00075 & 4.01703 & 0.9984 \\
        Zr0.01 Ce0.01 & 4.01088 & 4.00125 & 4.01890 & 0.9976 \\
        Zr0.01 Ce0.001 & 4.00302 & 3.99581 & 4.01885 & 0.9982 \\\hline
    \end{tabular} 
\end{table*}

The thermal hysteresis was calculated as 
$\Delta T = (A_s + A_f - M_s - M_f)/2$, in which the austenite start/finish $A_s$/$A_f$ and martensite start/finish $M_s$/$M_f$ temperatures were determined as the onsets of the heat emission/absorption peaks in Figure \ref{fig:ferro}(c). In our material system, the composition Zr0.005 Ce0.005 shows the smallest thermal hysteresis, 2.5$^\circ$C corresponding to the largest FOM among all tested compositions in Figure \ref{fig:corr}(a) and (b), meanwhile it satisfies closely the compatibility condition as indicated in Figure \ref{fig:corr}(c) corresponding to the structural parameters and compatibility indexers listed in Table \ref{tab:latt}. 
Our observation in tridoped BT system implies a strong correlation between the compatibility indexer $\lambda_2$ and
 the change of polarization across the first-order phase transformation. The maximum FOM resonates with the minimum of hysteresis corresponding to the most compatible first-order phase transformation.
The large FOM given by the composition Zr0.005 Ce0.005 makes it an ideal material candidate for the heat-to-electricity conversion demonstration.

\begin{figure}
    \centering
    \includegraphics[width=0.35\textwidth]{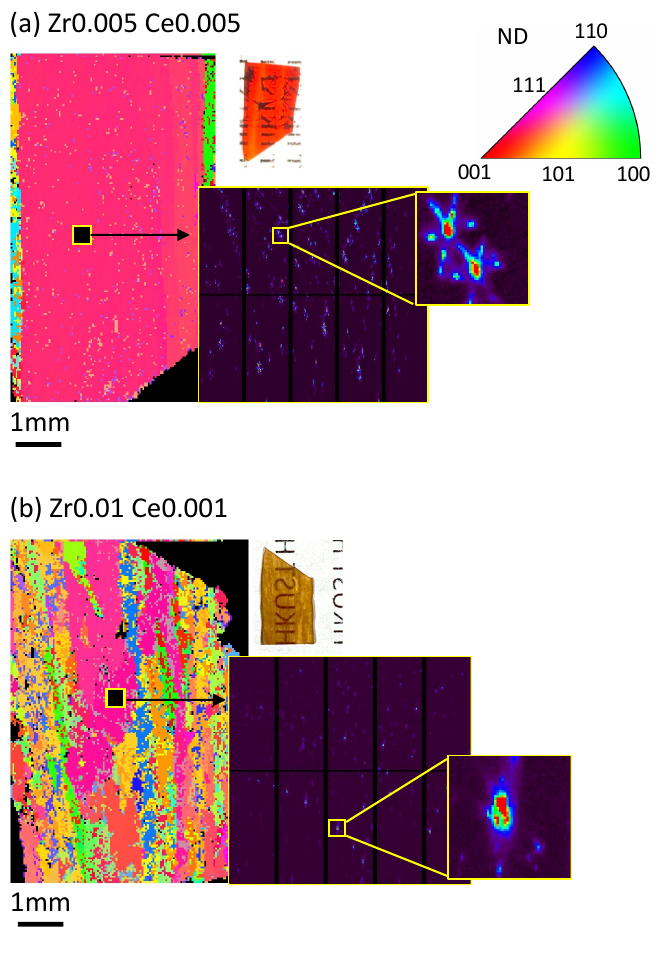}
    \caption{Orientation maps by synchrotron Laue microdiffraction for (a) Zr0.005 Ce0.005 (b) Zr0.01 Ce0.001 grown by floating zone method. The color map indicates the orientation of normal direction (ND) of the surface. }
    \label{fig:laue}
\end{figure}

\subsection{Steep polarization jump at phase transformation in single crystal}

For demonstration, it is critical to fabricate a high quality capacitor with the capacitance sensitive to temperature variation within the thermal fluctuation range. A previous energy conversion attempt \cite{bucsek2019} by first-order phase transformation evidently suggests that the single BaTiO$_3$ crystal can magnificently enhance the performance of the device.
We adopted the floating zone method, an economic way to grow single crystals, to coarsen the grains in two fine-grained polycrystals: Zr0.01 Ce0.001 and Zr0.005 Ce0.005. We selected two of the polycrystalline rods as the seed and feed aligned vertically for the crystal growth by floating zone method at a growth speed 10mm/hr and relative angular speed 25rpm at an increasing lamp power. The molten zone was created and monitored throughout the entire growth period. Finally, the as-grown rod was solidified as the furnace cools down. The rod was sliced along the growth direction followed by mechanical polish. The size of the ceramic slices is about $19.5\times 5.5\times 0.4$ mm$^3$. After mechanical polishing, both Zr0.01 Ce0.001 and Zr0.005 Ce0.005 slices are semi-transparent in the insets of Figure \ref{fig:laue}(a) and (b).

\begin{figure*}[htpb]
    \centering
    \includegraphics[width=0.65\textwidth]{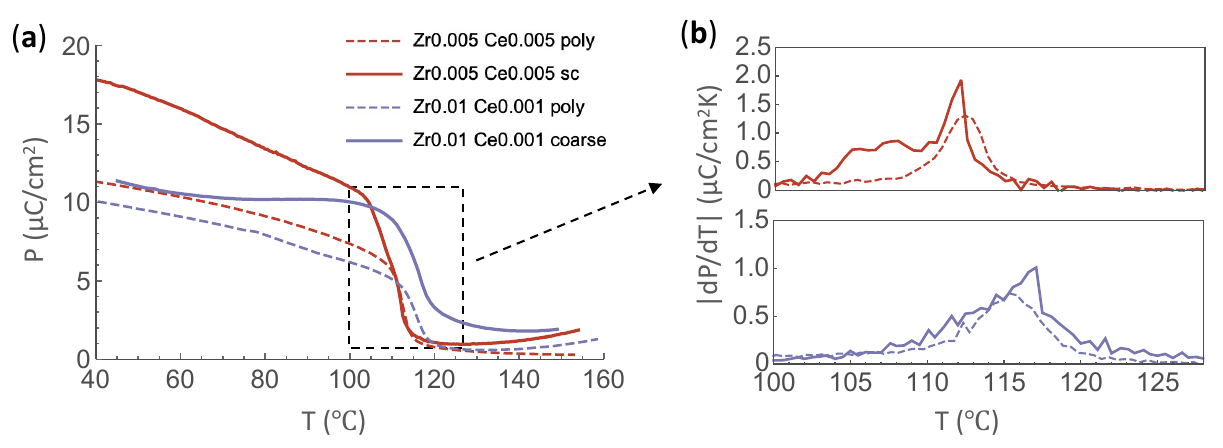}
    \caption{Ferroic properties of Zr0.005 Ce0.005 single crystal and Zr0.01 Ce0.001 coarse grain crystal. (a) Temperature dependent polarization and (b) absolute value of $dP/dT$ in a zoomed temperature range between 100 and 130$^\circ$C. }
    \label{fig:single}
\end{figure*}

Since we used polycrystal seeds for the floating zone process, the cutting process of the as-grown rod was done without acknowledgment of orientation information. 
Figure \ref{fig:laue} shows the post-cut orientation maps characterized by synchrotron X-ray Laue microdiffraction at Advanced Light Source, Beamline 12.3.2 at Lawrence Berkeley National Lab.
Clearly, the Zr0.005 Ce0.005 plate is a single crystal of millimeter size, while the Zr0.01 Ce0.001 plate is a coarse-grained crystal having multiple millimeter grains.
The surface normal of Zr0.005 Ce0.005 is measured as  [0.9, 1.5, 1.8] written in the crystal basis. Although it is not perfectly aligned with the polar axis [001], 
 its $\pjump$ reaches 11.7$\mu$C/cm$^2$, corresponding to $\kappa = 1.93\mu$C/cm$^{2}$K, shown in Figure \ref{fig:single}(a)-(b). 
 
 These values are about twice of the largest reported values in single crystal BaTiO$_3$ aligned with its polar axis \cite{moya2013}. 
Since the goal of energy conversion is to harvest low-grade waste heat, we consider 20$^\circ$C temperature difference near 100$^\circ$C to benchmark the ferroelectrics in our work and literature in Table \ref{tab:fom}. Zr0.005 Ce0.005 single crystal exhibits by far the largest polarization jump and pyroelectric coefficient.
 The coarsened Zr0.01 Ce0.001 crystal also exhibits some degree of improvement in $\pjump$ compared with its fine-grained polycrystal counterpart. 
Figure \ref{fig:single} compares the ferroic properties near the transformation temperature between the single/coarsened crystals and their polycrystalline form. First, we confirmed that both transform at the same temperature despite whether they are single or poly crystals. Second, both polarization jump and pyroelectric coefficient at transition temperature are magnified significantly. The leakage current densities are 0.064 $\mu$A/cm$^2$ for Zr0.005 Ce0.005, and 0.337 $\mu$A/cm$^2$ for Zr0.01 Ce0.001.

\subsection{Energy conversion demonstration for long thermal cycles}

\begin{figure}
    \centering
    \includegraphics[width=0.35\textwidth]{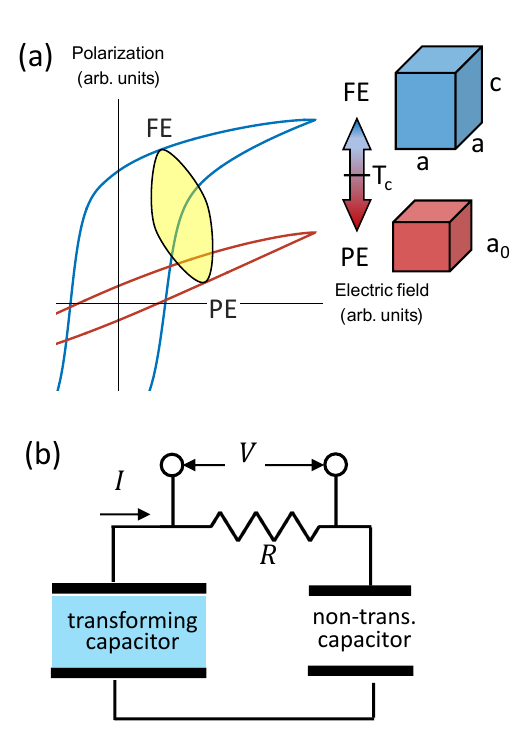}
    \caption{(a) Schematic polarization - field curves of tetragonal ferroelectric phase (blue) and cubic paraelectric phase (red). The yellow area demonstrates the thermodynamic cycle of the process without application of external bias electric field. (b) Battery detached setup of the transforming capacitor with polarization jump between ferroelectric and paraelectric phases. }
    \label{fig:setting}
\end{figure}

We make planar capacitors 
using the Zr0.005 Ce0.005 single crystal and the Zr0.01 Ce0.001 coarse-grained slices. The dimensions of the transforming capacitors are 16.5mm$^2\times 0.46$mm for Zr0.005 Ce0.005 single crystal capacitor and 25.28mm$^2\times$0.44mm for Zr0.01 Ce0.001 coarse-grain capacitor.
The demonstration apparatus is illustrated in Figure \ref{fig:setting}, which was analyzed by a thermodynamic model with complete removal of external bias field during energy conversion \cite{zhang2019power, zhang2020leakage}.

A non-transforming capacitor (10$\mu$F) used as charge reservoir is connected to the phase-transforming capacitor. Both are grounded and bridged via a resistor (1 M$\Omega$) to measure the electric signal generated by first-order phase transformation of the active material. 
The capacitors were initially charged up to hold sufficient initial charges, then detached from the external power source throughout the heating and cooling cycles. The heating/cooling rates mimic the natural thermal fluctuations from waste heat between 95 and 125$^\circ$C. The charges flow back and forth between the transforming and non-transforming capacitors driven by the ferro-to-paraelectric phase transformation.
 We measured the electric voltage $V$ by Agilent 34970A data acquisition module on the load resistor $R$ in Figure \ref{fig:setting}(b) to calculate the pyroelectric current as $V/R$, in short pyro-current\footnote{For simplicity, we still use the terminology \emph{pyro-current} as the current generated purely by phase transformation.}. 
 Note that the voltage signal can be magnified by using a large resistor, but it does not represent a true improvement of energy conversion performance,
 as the load resistance is given in practical applications.
 Hence, pyro-current is a more proper meter in our energy conversion setup.

\begin{figure*}
    \centering
    \includegraphics[width=0.7\textwidth]{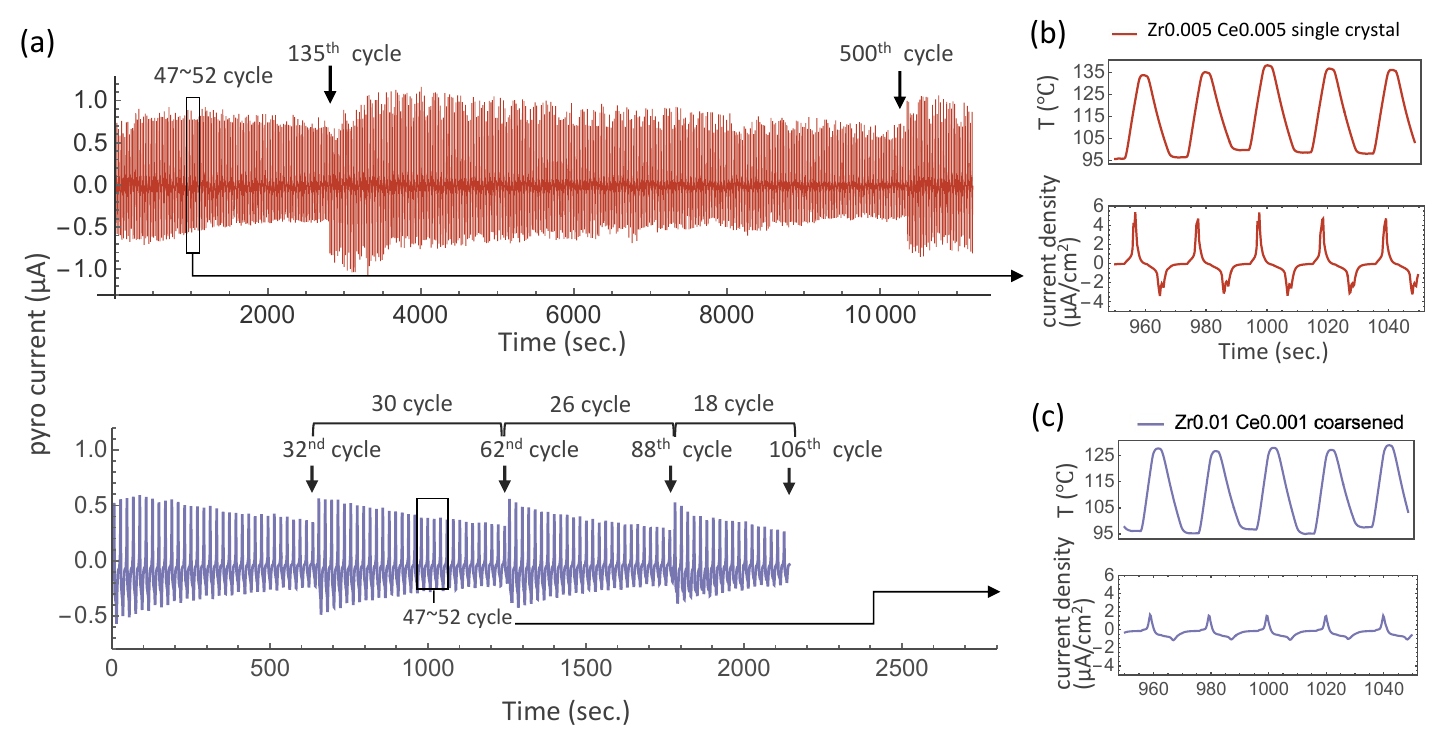}
    \caption{Energy conversion by phase-transforming ferroelectric crystals. (a) Pyro-current generated by the Zr0.005 Ce0.005 single crystal capacitor (red) and the Zr0.01 Ce0.001 coarse-grained capacitor (blue) over hundreds of transformation cycles. (b) - (c) are pyroelectric current density of 47 to 52 cycles corresponding to the applied temperature profiles. }
    \label{fig:demo}
\end{figure*}

Figure \ref{fig:demo}(a) demonstrates continuous generation of electricity more than 3 hours by Zr0.005 Ce0.005 capacitor, compared to the Zr0.01 Ce0.001 capacitor over hundreds of transformation cycles. The current is generated when temperature passes the phase-transformation temperatures during both heating and cooling processes, in Figure \ref{fig:demo}(b)-(c). It verifies that the electric energy is purely converted from temperature differences, not from the discharge of external battery.
The pyro-current generated by the Zr0.005 Ce0.005 single crystal capacitor reaches up to 1 $\mu$A stably and continuously lasting over 500 transformation cycles.
For the Zr0.01 Ce0.001 coarse-grained capacitor, the pyro-current generated in the first few cycles is 0.5 $\mu$A. In order to benchmark capacitors with different areas, we calculated the current density, i.e. current per area: 6 $\mu$A/cm$^2$ for Zr0.005 Ce0.005 and 2 $\mu$A/cm$^2$ for Zr0.01 Ce0.001. These results are aligned with our theoretical prediction by figure of merit for material development. This confirms that the FOM in eq. \eqref{eq:fom} is a good performance indexer of energy conversion by first-order phase transformation. 

As the Zr0.01 Ce0.001 capacitor exhibits larger electric leakage, it suffers electric degradation \cite{zhang2020leakage} over a period of energy conversion process. Consequently, a recharge is necessary after several thermodynamics cycles.    
The half-life -- number of cycles since a full recharge till the peak signal reduces 50\% -- shortens considerably after each full recharge.
It requires 4 full recharges to maintain a decent power output at the 106th cycle, labeled in the bottom panel of Figure \ref{fig:demo}(a). In contrast, the Zr0.005 Ce0.005 capacitor has no visible electric degradation during the first 135 cycles. After a recharge, the generation of pyro-current sustains for 365 cycles until the second recharge, corresponding to the top panel of Figure \ref{fig:demo}(a). An important observation is that the number of cycles in each half-life in the Zr0.01 Ce0.001 capacitor keeps shrinking, which implies a functional degradation of the transforming material.

\begin{figure}
    \centering
    \includegraphics[width=0.35\textwidth]{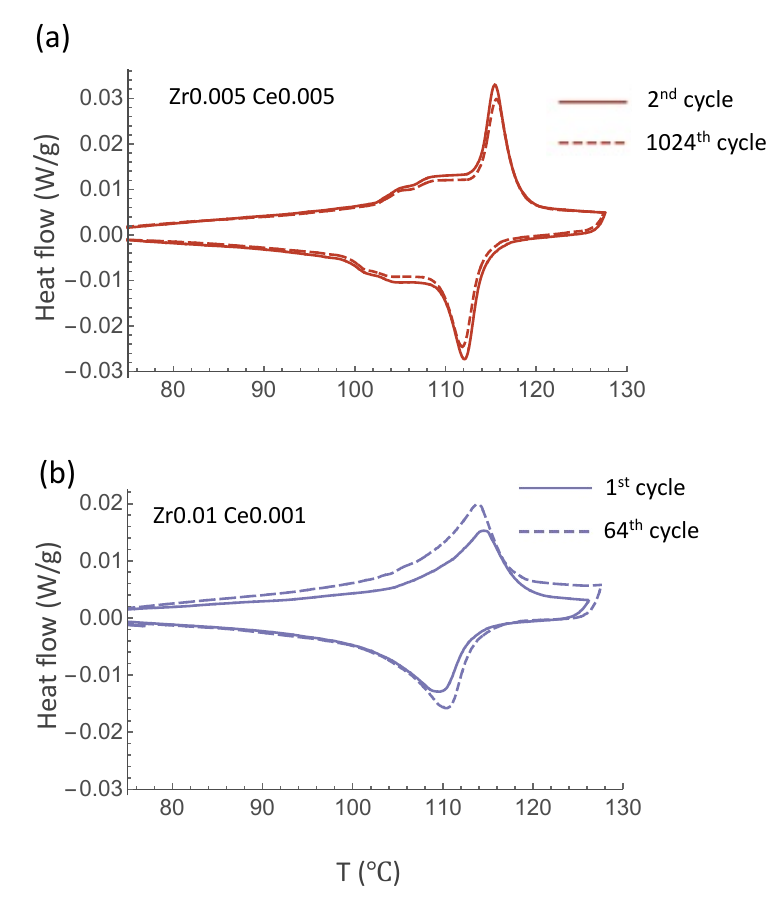}
    \caption{Differential scanning calorimetry measurement of (a) Zr0.005 Ce0.005 single crystal plate (red) and (b) Zr0.01 Ce0.001 coarse-grained plate (blue) at initial and final thermal cycles. }
    \label{fig:thermalcyc}
\end{figure}

The Zr0.005 Ce0.005 capacitor with $\lambda_2$ much closer to 1 exhibits magnificent reversibility for cyclic phase transformation. The thermal reversibility of these two capacitors were examined by cyclic DSC experiments, shown in Figure \ref{fig:thermalcyc}. The Zr0.005 Ce0.005 capacitor after 1024 thermal cycles shows no migration of transformation temperature (115$^\circ$C), nor the magnification of the hysteresis (2.7$^\circ$C) compared to the hysteresis in the first cycle (2.5$^\circ$C). This coincides with the satisfaction of the compatibility condition $\lambda_2=1$ (Figure \ref{fig:corr}b-c). 

\begin{figure}[ht]
    \centering
    \includegraphics[width=0.35\textwidth]{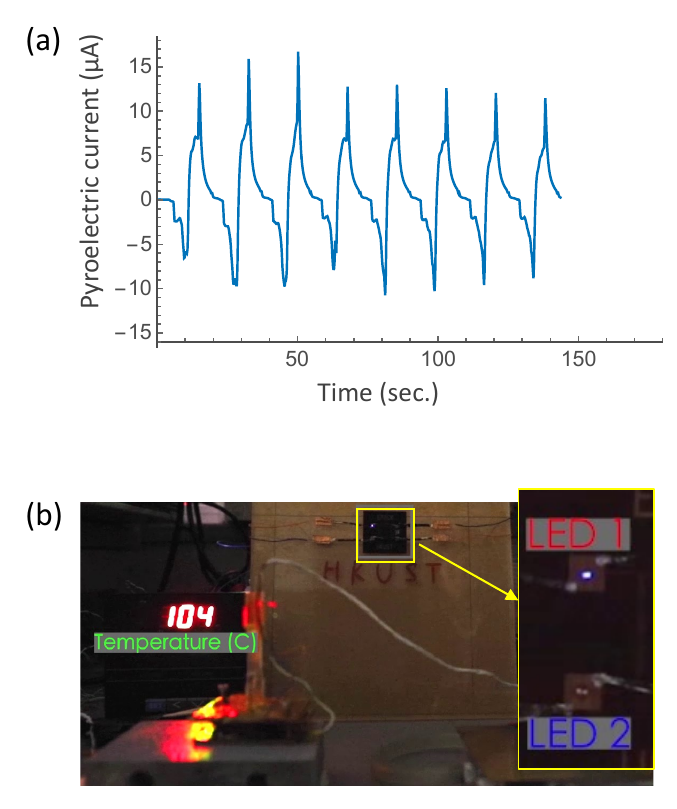}
    \caption{(a) Generated current by a grid of phase-transforming Zr0.005 Ce0.005 capacitors connected in parallel under $\pm 20^\circ$C temperature differences around 110$^\circ$C. (b) The demonstration to light up a LED light during heating. }
    \label{fig:38p}
\end{figure}

We then cut 38 pieces of the similar sizes from several rods of Zr0.005 Ce0.005 synthesized by the same method, and assembled them in parallel in our demonstration setup.  The total area of them is 2344.88mm$^2$, and the mean thickness is about 0.4mm. In this setting, we used a bigger reference capacitor (140$\mu$F) connected to the same load resistor. All capacitors were initially charged up to have 1.2 kV/cm electric field inside, then completely disconnect to the external power to begin the energy conversion cycles. The grid of energy conversion devices generates stably 15 $\mu$A electric current in consecutive cycles, in Figure \ref{fig:38p}(a). Thanks to the phase stability, low hysteresis and large pyro-current generated per cycle, the Zr0.005 Ce0.005 capacitor can directly light up a LED during both heating (Figure \ref{fig:38p}b) and cooling process through phase transformation (see Movie S1).

\section{Conclusions}

Enabling waste heat harvesting, particularly in relatively low temperature regime is an inter-disciplinary and multifaceted grand challenge in energy science.
Core material development is undoubtedly one of the key imperatives.
Rational figure of merit, phase compatibility and leakage suppression underlie essential design strategy for materials with minimal functional degradation and high energy conversion performance.
Although aspects like manageable temperature cycling, efficient heat transfer, collecting intermittent electricity and compact device packaging still call for breakthrough improvement, this work is a substantial step forward from the aspect of material development towards the practical realization of pyroelectric energy conversion.




\section*{Acknowledgments}

The authors appreciate fruitful discussion with Dr. Yintao Song during the preparation of the manuscript. Lunhua He and Sihao Deng are acknowledged for their help on the neutron powder diffraction experiments which were performed at GPPD of the China Spallation Neutron Source (CSNS), Dongguan, China.
 C. Z., Z. Zeng, Z. Zhu and X. C. are grateful for financial support under GRF grants 16201118 and 16201019, CRF grant C6016-20G  from the HK Research Grants Council  and the Bridge Gap Fund, BGF.009.19/20, from the HKUST Technology Transfer
Center. This research used BL12.3.2 which is a resource of the Advanced Light Source, a U.S. DOE Office of Science User Facility under contract no. DE-AC02-05CH11231.

\bibliography{ref_giant_pyro}

\begin{thebibliography}{49}%
\makeatletter
\providecommand \@ifxundefined [1]{%
 \@ifx{#1\undefined}
}%
\providecommand \@ifnum [1]{%
 \ifnum #1\expandafter \@firstoftwo
 \else \expandafter \@secondoftwo
 \fi
}%
\providecommand \@ifx [1]{%
 \ifx #1\expandafter \@firstoftwo
 \else \expandafter \@secondoftwo
 \fi
}%
\providecommand \natexlab [1]{#1}%
\providecommand \enquote  [1]{``#1''}%
\providecommand \bibnamefont  [1]{#1}%
\providecommand \bibfnamefont [1]{#1}%
\providecommand \citenamefont [1]{#1}%
\providecommand \href@noop [0]{\@secondoftwo}%
\providecommand \href [0]{\begingroup \@sanitize@url \@href}%
\providecommand \@href[1]{\@@startlink{#1}\@@href}%
\providecommand \@@href[1]{\endgroup#1\@@endlink}%
\providecommand \@sanitize@url [0]{\catcode `\\12\catcode `\$12\catcode
  `\&12\catcode `\#12\catcode `\^12\catcode `\_12\catcode `\%12\relax}%
\providecommand \@@startlink[1]{}%
\providecommand \@@endlink[0]{}%
\providecommand \url  [0]{\begingroup\@sanitize@url \@url }%
\providecommand \@url [1]{\endgroup\@href {#1}{\urlprefix }}%
\providecommand \urlprefix  [0]{URL }%
\providecommand \Eprint [0]{\href }%
\providecommand \doibase [0]{https://doi.org/}%
\providecommand \selectlanguage [0]{\@gobble}%
\providecommand \bibinfo  [0]{\@secondoftwo}%
\providecommand \bibfield  [0]{\@secondoftwo}%
\providecommand \translation [1]{[#1]}%
\providecommand \BibitemOpen [0]{}%
\providecommand \bibitemStop [0]{}%
\providecommand \bibitemNoStop [0]{.\EOS\space}%
\providecommand \EOS [0]{\spacefactor3000\relax}%
\providecommand \BibitemShut  [1]{\csname bibitem#1\endcsname}%
\let\auto@bib@innerbib\@empty
\bibitem [{\citenamefont {Lang}(1974)}]{lang1974}%
  \BibitemOpen
  \bibfield  {author} {\bibinfo {author} {\bibfnamefont {S.~B.}\ \bibnamefont
  {Lang}},\ }\bibfield  {title} {\bibinfo {title} {Pyroelectricity: A 2300-year
  history},\ }\href@noop {} {\bibfield  {journal} {\bibinfo  {journal}
  {Ferroelectrics}\ }\textbf {\bibinfo {volume} {7}},\ \bibinfo {pages} {231}
  (\bibinfo {year} {1974})}\BibitemShut {NoStop}%
\bibitem [{\citenamefont {Whatmore}(1986)}]{whatmore1986}%
  \BibitemOpen
  \bibfield  {author} {\bibinfo {author} {\bibfnamefont {R.}~\bibnamefont
  {Whatmore}},\ }\bibfield  {title} {\bibinfo {title} {Pyroelectric devices and
  materials},\ }\href@noop {} {\bibfield  {journal} {\bibinfo  {journal} {Rep.
  Prog. Phys.}\ }\textbf {\bibinfo {volume} {49}},\ \bibinfo {pages} {1335}
  (\bibinfo {year} {1986})}\BibitemShut {NoStop}%
\bibitem [{\citenamefont {Bauer}\ and\ \citenamefont
  {Ploss}(1989)}]{bauer1989simple}%
  \BibitemOpen
  \bibfield  {author} {\bibinfo {author} {\bibfnamefont {S.}~\bibnamefont
  {Bauer}}\ and\ \bibinfo {author} {\bibfnamefont {B.}~\bibnamefont {Ploss}},\
  }\bibfield  {title} {\bibinfo {title} {A simple technique to interface
  pyroelectric materials with silicon substrates for infrared detection},\
  }\href@noop {} {\bibfield  {journal} {\bibinfo  {journal} {Ferroelectrics
  Letters Section}\ }\textbf {\bibinfo {volume} {9}},\ \bibinfo {pages} {155}
  (\bibinfo {year} {1989})}\BibitemShut {NoStop}%
\bibitem [{\citenamefont {Drummond}\ \emph {et~al.}(1979)\citenamefont
  {Drummond}, \citenamefont {Fargo}, \citenamefont {Ream}, \citenamefont
  {Briscoe},\ and\ \citenamefont {Brown}}]{drummond1979}%
  \BibitemOpen
  \bibfield  {author} {\bibinfo {author} {\bibfnamefont {J.}~\bibnamefont
  {Drummond}}, \bibinfo {author} {\bibfnamefont {V.}~\bibnamefont {Fargo}},
  \bibinfo {author} {\bibfnamefont {J.}~\bibnamefont {Ream}}, \bibinfo {author}
  {\bibfnamefont {J.}~\bibnamefont {Briscoe}},\ and\ \bibinfo {author}
  {\bibfnamefont {D.}~\bibnamefont {Brown}},\ }\bibfield  {title} {\bibinfo
  {title} {Experimental demonstration of heat-to-electricity conversion within
  a dielectric},\ }in\ \href@noop {} {\emph {\bibinfo {booktitle} {14th
  Intersociety Energy Conversion Engineering Conference}}},\ Vol.~\bibinfo
  {volume} {2}\ (\bibinfo {year} {1979})\ pp.\ \bibinfo {pages}
  {1819--1824}\BibitemShut {NoStop}%
\bibitem [{\citenamefont {Olsen}\ and\ \citenamefont
  {Evans}(1983)}]{olsen1983}%
  \BibitemOpen
  \bibfield  {author} {\bibinfo {author} {\bibfnamefont {R.~B.}\ \bibnamefont
  {Olsen}}\ and\ \bibinfo {author} {\bibfnamefont {D.}~\bibnamefont {Evans}},\
  }\bibfield  {title} {\bibinfo {title} {Pyroelectric energy conversion:
  hysteresis loss and temperature sensitivity of a ferroelectric material},\
  }\href@noop {} {\bibfield  {journal} {\bibinfo  {journal} {J. Appl. Phys.}\
  }\textbf {\bibinfo {volume} {54}},\ \bibinfo {pages} {5941} (\bibinfo {year}
  {1983})}\BibitemShut {NoStop}%
\bibitem [{\citenamefont {Olsen}\ \emph {et~al.}(1985)\citenamefont {Olsen},
  \citenamefont {Bruno},\ and\ \citenamefont {Briscoe}}]{olsen1985cycle}%
  \BibitemOpen
  \bibfield  {author} {\bibinfo {author} {\bibfnamefont {R.~B.}\ \bibnamefont
  {Olsen}}, \bibinfo {author} {\bibfnamefont {D.~A.}\ \bibnamefont {Bruno}},\
  and\ \bibinfo {author} {\bibfnamefont {J.~M.}\ \bibnamefont {Briscoe}},\
  }\bibfield  {title} {\bibinfo {title} {Pyroelectric conversion cycles},\
  }\href@noop {} {\bibfield  {journal} {\bibinfo  {journal} {J. Appl. Phys.}\
  }\textbf {\bibinfo {volume} {58}},\ \bibinfo {pages} {4709} (\bibinfo {year}
  {1985})}\BibitemShut {NoStop}%
\bibitem [{\citenamefont {Sebald}\ \emph
  {et~al.}(2008{\natexlab{a}})\citenamefont {Sebald}, \citenamefont {Pruvost},\
  and\ \citenamefont {Guyomar}}]{sebald2007}%
  \BibitemOpen
  \bibfield  {author} {\bibinfo {author} {\bibfnamefont {G.}~\bibnamefont
  {Sebald}}, \bibinfo {author} {\bibfnamefont {S.}~\bibnamefont {Pruvost}},\
  and\ \bibinfo {author} {\bibfnamefont {D.}~\bibnamefont {Guyomar}},\
  }\bibfield  {title} {\bibinfo {title} {Energy harvesting based on ericsson
  pyroelectric cycles in a relaxor ferroelectric ceramic},\ }\href@noop {}
  {\bibfield  {journal} {\bibinfo  {journal} {Smart Mater. Struct.}\ }\textbf
  {\bibinfo {volume} {17}},\ \bibinfo {pages} {015012} (\bibinfo {year}
  {2008}{\natexlab{a}})}\BibitemShut {NoStop}%
\bibitem [{\citenamefont {Sebald}\ \emph {et~al.}(2009)\citenamefont {Sebald},
  \citenamefont {Guyomar},\ and\ \citenamefont {Agbossou}}]{sebald2009}%
  \BibitemOpen
  \bibfield  {author} {\bibinfo {author} {\bibfnamefont {G.}~\bibnamefont
  {Sebald}}, \bibinfo {author} {\bibfnamefont {D.}~\bibnamefont {Guyomar}},\
  and\ \bibinfo {author} {\bibfnamefont {A.}~\bibnamefont {Agbossou}},\
  }\bibfield  {title} {\bibinfo {title} {On thermoelectric and pyroelectric
  energy harvesting},\ }\href@noop {} {\bibfield  {journal} {\bibinfo
  {journal} {Smart Mater. Struct.}\ }\textbf {\bibinfo {volume} {18}},\
  \bibinfo {pages} {125006} (\bibinfo {year} {2009})}\BibitemShut {NoStop}%
\bibitem [{\citenamefont {Navid}\ and\ \citenamefont
  {Pilon}(2011)}]{navid2011}%
  \BibitemOpen
  \bibfield  {author} {\bibinfo {author} {\bibfnamefont {A.}~\bibnamefont
  {Navid}}\ and\ \bibinfo {author} {\bibfnamefont {L.}~\bibnamefont {Pilon}},\
  }\bibfield  {title} {\bibinfo {title} {Pyroelectric energy harvesting using
  {Olsen} cycles in purified and porous poly (vinylidene
  fluoride-trifluoroethylene)[{P (VDF-TrFE)}] thin films},\ }\href@noop {}
  {\bibfield  {journal} {\bibinfo  {journal} {Smart Mater. Struct.}\ }\textbf
  {\bibinfo {volume} {20}},\ \bibinfo {pages} {025012} (\bibinfo {year}
  {2011})}\BibitemShut {NoStop}%
\bibitem [{\citenamefont {Yang}\ \emph {et~al.}(2012)\citenamefont {Yang},
  \citenamefont {Wang}, \citenamefont {Zhang},\ and\ \citenamefont
  {Wang}}]{yang2012}%
  \BibitemOpen
  \bibfield  {author} {\bibinfo {author} {\bibfnamefont {Y.}~\bibnamefont
  {Yang}}, \bibinfo {author} {\bibfnamefont {S.}~\bibnamefont {Wang}}, \bibinfo
  {author} {\bibfnamefont {Y.}~\bibnamefont {Zhang}},\ and\ \bibinfo {author}
  {\bibfnamefont {Z.~L.}\ \bibnamefont {Wang}},\ }\bibfield  {title} {\bibinfo
  {title} {Pyroelectric nanogenerators for driving wireless sensors},\
  }\href@noop {} {\bibfield  {journal} {\bibinfo  {journal} {Nano Lett.}\
  }\textbf {\bibinfo {volume} {12}},\ \bibinfo {pages} {6408} (\bibinfo {year}
  {2012})}\BibitemShut {NoStop}%
\bibitem [{\citenamefont {Bhatia}\ \emph {et~al.}(2014)\citenamefont {Bhatia},
  \citenamefont {Damodaran}, \citenamefont {Cho}, \citenamefont {Martin},\ and\
  \citenamefont {King}}]{bhatia2014high}%
  \BibitemOpen
  \bibfield  {author} {\bibinfo {author} {\bibfnamefont {B.}~\bibnamefont
  {Bhatia}}, \bibinfo {author} {\bibfnamefont {A.~R.}\ \bibnamefont
  {Damodaran}}, \bibinfo {author} {\bibfnamefont {H.}~\bibnamefont {Cho}},
  \bibinfo {author} {\bibfnamefont {L.~W.}\ \bibnamefont {Martin}},\ and\
  \bibinfo {author} {\bibfnamefont {W.~P.}\ \bibnamefont {King}},\ }\bibfield
  {title} {\bibinfo {title} {High-frequency thermal-electrical cycles for
  pyroelectric energy conversion},\ }\href@noop {} {\bibfield  {journal}
  {\bibinfo  {journal} {J. Appl. Phys.}\ }\textbf {\bibinfo {volume} {116}},\
  \bibinfo {pages} {194509} (\bibinfo {year} {2014})}\BibitemShut {NoStop}%
\bibitem [{\citenamefont {Pandya}\ \emph {et~al.}(2018)\citenamefont {Pandya},
  \citenamefont {Wilbur}, \citenamefont {Kim}, \citenamefont {Gao},
  \citenamefont {Dasgupta}, \citenamefont {Dames},\ and\ \citenamefont
  {Martin}}]{pandya2018nm}%
  \BibitemOpen
  \bibfield  {author} {\bibinfo {author} {\bibfnamefont {S.}~\bibnamefont
  {Pandya}}, \bibinfo {author} {\bibfnamefont {J.}~\bibnamefont {Wilbur}},
  \bibinfo {author} {\bibfnamefont {J.}~\bibnamefont {Kim}}, \bibinfo {author}
  {\bibfnamefont {R.}~\bibnamefont {Gao}}, \bibinfo {author} {\bibfnamefont
  {A.}~\bibnamefont {Dasgupta}}, \bibinfo {author} {\bibfnamefont
  {C.}~\bibnamefont {Dames}},\ and\ \bibinfo {author} {\bibfnamefont {L.~W.}\
  \bibnamefont {Martin}},\ }\bibfield  {title} {\bibinfo {title} {Pyroelectric
  energy conversion with large energy and power density in relaxor
  ferroelectric thin films},\ }\href@noop {} {\bibfield  {journal} {\bibinfo
  {journal} {Nat. Mater.}\ }\textbf {\bibinfo {volume} {17}},\ \bibinfo {pages}
  {432} (\bibinfo {year} {2018})}\BibitemShut {NoStop}%
\bibitem [{\citenamefont {Sharma}\ \emph {et~al.}(2021)\citenamefont {Sharma},
  \citenamefont {Behera}, \citenamefont {Pradhan}, \citenamefont {Pradhan},
  \citenamefont {Bonner},\ and\ \citenamefont {Bahoura}}]{sharma2021lead}%
  \BibitemOpen
  \bibfield  {author} {\bibinfo {author} {\bibfnamefont {A.~P.}\ \bibnamefont
  {Sharma}}, \bibinfo {author} {\bibfnamefont {M.~K.}\ \bibnamefont {Behera}},
  \bibinfo {author} {\bibfnamefont {D.~K.}\ \bibnamefont {Pradhan}}, \bibinfo
  {author} {\bibfnamefont {S.~K.}\ \bibnamefont {Pradhan}}, \bibinfo {author}
  {\bibfnamefont {C.~E.}\ \bibnamefont {Bonner}},\ and\ \bibinfo {author}
  {\bibfnamefont {M.}~\bibnamefont {Bahoura}},\ }\bibfield  {title} {\bibinfo
  {title} {Lead-free relaxor-ferroelectric thin films for energy harvesting
  from low-grade waste-heat},\ }\href@noop {} {\bibfield  {journal} {\bibinfo
  {journal} {Scientific Reports}\ }\textbf {\bibinfo {volume} {11}},\ \bibinfo
  {pages} {1} (\bibinfo {year} {2021})}\BibitemShut {NoStop}%
\bibitem [{\citenamefont {Pandya}\ \emph {et~al.}(2017)\citenamefont {Pandya},
  \citenamefont {Wilbur}, \citenamefont {Bhatia}, \citenamefont {Damodaran},
  \citenamefont {Monachon}, \citenamefont {Dasgupta}, \citenamefont {King},
  \citenamefont {Dames},\ and\ \citenamefont {Martin}}]{pandya2017prapplied}%
  \BibitemOpen
  \bibfield  {author} {\bibinfo {author} {\bibfnamefont {S.}~\bibnamefont
  {Pandya}}, \bibinfo {author} {\bibfnamefont {J.~D.}\ \bibnamefont {Wilbur}},
  \bibinfo {author} {\bibfnamefont {B.}~\bibnamefont {Bhatia}}, \bibinfo
  {author} {\bibfnamefont {A.~R.}\ \bibnamefont {Damodaran}}, \bibinfo {author}
  {\bibfnamefont {C.}~\bibnamefont {Monachon}}, \bibinfo {author}
  {\bibfnamefont {A.}~\bibnamefont {Dasgupta}}, \bibinfo {author}
  {\bibfnamefont {W.~P.}\ \bibnamefont {King}}, \bibinfo {author}
  {\bibfnamefont {C.}~\bibnamefont {Dames}},\ and\ \bibinfo {author}
  {\bibfnamefont {L.~W.}\ \bibnamefont {Martin}},\ }\bibfield  {title}
  {\bibinfo {title} {Direct measurement of pyroelectric and electrocaloric
  effects in thin films},\ }\href@noop {} {\bibfield  {journal} {\bibinfo
  {journal} {Phys. Rev. Applied}\ }\textbf {\bibinfo {volume} {7}},\ \bibinfo
  {pages} {034025} (\bibinfo {year} {2017})}\BibitemShut {NoStop}%
\bibitem [{\citenamefont {Ji}\ \emph {et~al.}(2019)\citenamefont {Ji},
  \citenamefont {Zhang}, \citenamefont {Wang},\ and\ \citenamefont
  {Yang}}]{ji2019piezo}%
  \BibitemOpen
  \bibfield  {author} {\bibinfo {author} {\bibfnamefont {Y.}~\bibnamefont
  {Ji}}, \bibinfo {author} {\bibfnamefont {K.}~\bibnamefont {Zhang}}, \bibinfo
  {author} {\bibfnamefont {Z.~L.}\ \bibnamefont {Wang}},\ and\ \bibinfo
  {author} {\bibfnamefont {Y.}~\bibnamefont {Yang}},\ }\bibfield  {title}
  {\bibinfo {title} {Piezo--pyro--photoelectric effects induced coupling
  enhancement of charge quantity in {BaTiO3} materials for simultaneously
  scavenging light and vibration energies},\ }\href@noop {} {\bibfield
  {journal} {\bibinfo  {journal} {Energy Environ. Sci.}\ }\textbf {\bibinfo
  {volume} {12}},\ \bibinfo {pages} {1231} (\bibinfo {year}
  {2019})}\BibitemShut {NoStop}%
\bibitem [{\citenamefont {Moya}\ \emph {et~al.}(2013)\citenamefont {Moya},
  \citenamefont {Stern-Taulats}, \citenamefont {Crossley}, \citenamefont
  {Gonz{\'a}lez-Alonso}, \citenamefont {Kar-Narayan}, \citenamefont {Planes},
  \citenamefont {Ma{\~n}osa},\ and\ \citenamefont {Mathur}}]{moya2013}%
  \BibitemOpen
  \bibfield  {author} {\bibinfo {author} {\bibfnamefont {X.}~\bibnamefont
  {Moya}}, \bibinfo {author} {\bibfnamefont {E.}~\bibnamefont {Stern-Taulats}},
  \bibinfo {author} {\bibfnamefont {S.}~\bibnamefont {Crossley}}, \bibinfo
  {author} {\bibfnamefont {D.}~\bibnamefont {Gonz{\'a}lez-Alonso}}, \bibinfo
  {author} {\bibfnamefont {S.}~\bibnamefont {Kar-Narayan}}, \bibinfo {author}
  {\bibfnamefont {A.}~\bibnamefont {Planes}}, \bibinfo {author} {\bibfnamefont
  {L.}~\bibnamefont {Ma{\~n}osa}},\ and\ \bibinfo {author} {\bibfnamefont
  {N.~D.}\ \bibnamefont {Mathur}},\ }\bibfield  {title} {\bibinfo {title}
  {Giant electrocaloric strength in single-crystal {BaTiO3}},\ }\href@noop {}
  {\bibfield  {journal} {\bibinfo  {journal} {Adv. Mater.}\ }\textbf {\bibinfo
  {volume} {25}},\ \bibinfo {pages} {1360} (\bibinfo {year}
  {2013})}\BibitemShut {NoStop}%
\bibitem [{\citenamefont {Kumar}\ \emph {et~al.}(2004)\citenamefont {Kumar},
  \citenamefont {Sharma}, \citenamefont {Thakur}, \citenamefont {Prakash},\
  and\ \citenamefont {Goel}}]{kumar2004}%
  \BibitemOpen
  \bibfield  {author} {\bibinfo {author} {\bibfnamefont {P.}~\bibnamefont
  {Kumar}}, \bibinfo {author} {\bibfnamefont {S.}~\bibnamefont {Sharma}},
  \bibinfo {author} {\bibfnamefont {O.}~\bibnamefont {Thakur}}, \bibinfo
  {author} {\bibfnamefont {C.}~\bibnamefont {Prakash}},\ and\ \bibinfo {author}
  {\bibfnamefont {T.}~\bibnamefont {Goel}},\ }\bibfield  {title} {\bibinfo
  {title} {Dielectric, piezoelectric and pyroelectric properties of {PMN--PT}
  (68: 32) system},\ }\href@noop {} {\bibfield  {journal} {\bibinfo  {journal}
  {Ceramics International}\ }\textbf {\bibinfo {volume} {30}},\ \bibinfo
  {pages} {585} (\bibinfo {year} {2004})}\BibitemShut {NoStop}%
\bibitem [{\citenamefont {Bucsek}\ \emph {et~al.}(2019)\citenamefont {Bucsek},
  \citenamefont {Nunn}, \citenamefont {Jalan},\ and\ \citenamefont
  {James}}]{bucsek2019}%
  \BibitemOpen
  \bibfield  {author} {\bibinfo {author} {\bibfnamefont {A.}~\bibnamefont
  {Bucsek}}, \bibinfo {author} {\bibfnamefont {W.}~\bibnamefont {Nunn}},
  \bibinfo {author} {\bibfnamefont {B.}~\bibnamefont {Jalan}},\ and\ \bibinfo
  {author} {\bibfnamefont {R.~D.}\ \bibnamefont {James}},\ }\bibfield  {title}
  {\bibinfo {title} {Direct conversion of heat to electricity using first-order
  phase transformations in ferroelectrics},\ }\href@noop {} {\bibfield
  {journal} {\bibinfo  {journal} {Phys. Rev. Applied}\ }\textbf {\bibinfo
  {volume} {12}},\ \bibinfo {pages} {034043} (\bibinfo {year}
  {2019})}\BibitemShut {NoStop}%
\bibitem [{\citenamefont {Zhang}\ \emph
  {et~al.}(2020{\natexlab{a}})\citenamefont {Zhang}, \citenamefont {Phuong},
  \citenamefont {Roake}, \citenamefont {Khanbareh}, \citenamefont {Wang},
  \citenamefont {Dunn},\ and\ \citenamefont {Bowen}}]{zhang2020thermal}%
  \BibitemOpen
  \bibfield  {author} {\bibinfo {author} {\bibfnamefont {Y.}~\bibnamefont
  {Zhang}}, \bibinfo {author} {\bibfnamefont {P.~T.~T.}\ \bibnamefont
  {Phuong}}, \bibinfo {author} {\bibfnamefont {E.}~\bibnamefont {Roake}},
  \bibinfo {author} {\bibfnamefont {H.}~\bibnamefont {Khanbareh}}, \bibinfo
  {author} {\bibfnamefont {Y.}~\bibnamefont {Wang}}, \bibinfo {author}
  {\bibfnamefont {S.}~\bibnamefont {Dunn}},\ and\ \bibinfo {author}
  {\bibfnamefont {C.}~\bibnamefont {Bowen}},\ }\bibfield  {title} {\bibinfo
  {title} {Thermal energy harvesting using pyroelectric-electrochemical
  coupling in ferroelectric materials},\ }\href@noop {} {\bibfield  {journal}
  {\bibinfo  {journal} {Joule}\ }\textbf {\bibinfo {volume} {4}},\ \bibinfo
  {pages} {301} (\bibinfo {year} {2020}{\natexlab{a}})}\BibitemShut {NoStop}%
\bibitem [{\citenamefont {Lang}\ and\ \citenamefont
  {Das-Gupta}(2001)}]{lang2001book}%
  \BibitemOpen
  \bibfield  {author} {\bibinfo {author} {\bibfnamefont {S.~B.}\ \bibnamefont
  {Lang}}\ and\ \bibinfo {author} {\bibfnamefont {D.~K.}\ \bibnamefont
  {Das-Gupta}},\ }\bibfield  {title} {\bibinfo {title} {Pyroelectricity:
  Fundamentals and applications},\ }in\ \href@noop {} {\emph {\bibinfo
  {booktitle} {Handbook of advanced electronic and photonic materials and
  devices}}}\ (\bibinfo  {publisher} {Elsevier},\ \bibinfo {year} {2001})\ pp.\
  \bibinfo {pages} {1--55}\BibitemShut {NoStop}%
\bibitem [{\citenamefont {Bowen}\ \emph {et~al.}(2014)\citenamefont {Bowen},
  \citenamefont {Taylor}, \citenamefont {LeBoulbar}, \citenamefont {Zabek},
  \citenamefont {Chauhan},\ and\ \citenamefont {Vaish}}]{bowen2014}%
  \BibitemOpen
  \bibfield  {author} {\bibinfo {author} {\bibfnamefont {C.~R.}\ \bibnamefont
  {Bowen}}, \bibinfo {author} {\bibfnamefont {J.}~\bibnamefont {Taylor}},
  \bibinfo {author} {\bibfnamefont {E.}~\bibnamefont {LeBoulbar}}, \bibinfo
  {author} {\bibfnamefont {D.}~\bibnamefont {Zabek}}, \bibinfo {author}
  {\bibfnamefont {A.}~\bibnamefont {Chauhan}},\ and\ \bibinfo {author}
  {\bibfnamefont {R.}~\bibnamefont {Vaish}},\ }\bibfield  {title} {\bibinfo
  {title} {Pyroelectric materials and devices for energy harvesting
  applications},\ }\href@noop {} {\bibfield  {journal} {\bibinfo  {journal}
  {Energy Environ. Sci.}\ }\textbf {\bibinfo {volume} {7}},\ \bibinfo {pages}
  {3836} (\bibinfo {year} {2014})}\BibitemShut {NoStop}%
\bibitem [{\citenamefont {Sebald}\ \emph {et~al.}(2006)\citenamefont {Sebald},
  \citenamefont {Seveyrat}, \citenamefont {Guyomar}, \citenamefont {Lebrun},
  \citenamefont {Guiffard},\ and\ \citenamefont {Pruvost}}]{sebald2006fom}%
  \BibitemOpen
  \bibfield  {author} {\bibinfo {author} {\bibfnamefont {G.}~\bibnamefont
  {Sebald}}, \bibinfo {author} {\bibfnamefont {L.}~\bibnamefont {Seveyrat}},
  \bibinfo {author} {\bibfnamefont {D.}~\bibnamefont {Guyomar}}, \bibinfo
  {author} {\bibfnamefont {L.}~\bibnamefont {Lebrun}}, \bibinfo {author}
  {\bibfnamefont {B.}~\bibnamefont {Guiffard}},\ and\ \bibinfo {author}
  {\bibfnamefont {S.}~\bibnamefont {Pruvost}},\ }\bibfield  {title} {\bibinfo
  {title} {Electrocaloric and pyroelectric properties of {0.75Pb(Mg1/3
  Nb2/3)O$_3$--0.25PbTiO$_3$} single crystals},\ }\href@noop {} {\bibfield
  {journal} {\bibinfo  {journal} {Journal of applied physics}\ }\textbf
  {\bibinfo {volume} {100}},\ \bibinfo {pages} {124112} (\bibinfo {year}
  {2006})}\BibitemShut {NoStop}%
\bibitem [{\citenamefont {Sebald}\ \emph
  {et~al.}(2008{\natexlab{b}})\citenamefont {Sebald}, \citenamefont
  {Lefeuvre},\ and\ \citenamefont {Guyomar}}]{sebald2008pyroelectric}%
  \BibitemOpen
  \bibfield  {author} {\bibinfo {author} {\bibfnamefont {G.}~\bibnamefont
  {Sebald}}, \bibinfo {author} {\bibfnamefont {E.}~\bibnamefont {Lefeuvre}},\
  and\ \bibinfo {author} {\bibfnamefont {D.}~\bibnamefont {Guyomar}},\
  }\bibfield  {title} {\bibinfo {title} {Pyroelectric energy conversion:
  optimization principles},\ }\href@noop {} {\bibfield  {journal} {\bibinfo
  {journal} {IEEE transactions on ultrasonics, ferroelectrics, and frequency
  control}\ }\textbf {\bibinfo {volume} {55}},\ \bibinfo {pages} {538}
  (\bibinfo {year} {2008}{\natexlab{b}})}\BibitemShut {NoStop}%
\bibitem [{\citenamefont {Mangalam}\ \emph {et~al.}(2013)\citenamefont
  {Mangalam}, \citenamefont {Agar}, \citenamefont {Damodaran}, \citenamefont
  {Karthik},\ and\ \citenamefont {Martin}}]{mangalam2013improved}%
  \BibitemOpen
  \bibfield  {author} {\bibinfo {author} {\bibfnamefont {R.}~\bibnamefont
  {Mangalam}}, \bibinfo {author} {\bibfnamefont {J.}~\bibnamefont {Agar}},
  \bibinfo {author} {\bibfnamefont {A.}~\bibnamefont {Damodaran}}, \bibinfo
  {author} {\bibfnamefont {J.}~\bibnamefont {Karthik}},\ and\ \bibinfo {author}
  {\bibfnamefont {L.}~\bibnamefont {Martin}},\ }\bibfield  {title} {\bibinfo
  {title} {Improved pyroelectric figures of merit in compositionally graded
  {PbZr$_{1-x}$Ti$_x$O$_3$} thin films},\ }\href@noop {} {\bibfield  {journal}
  {\bibinfo  {journal} {ACS applied materials \& interfaces}\ }\textbf
  {\bibinfo {volume} {5}},\ \bibinfo {pages} {13235} (\bibinfo {year}
  {2013})}\BibitemShut {NoStop}%
\bibitem [{\citenamefont {Zhang}\ \emph {et~al.}(2019)\citenamefont {Zhang},
  \citenamefont {Song}, \citenamefont {Wegner}, \citenamefont {Quandt},
  \citenamefont {Chen} \emph {et~al.}}]{zhang2019power}%
  \BibitemOpen
  \bibfield  {author} {\bibinfo {author} {\bibfnamefont {C.}~\bibnamefont
  {Zhang}}, \bibinfo {author} {\bibfnamefont {Y.}~\bibnamefont {Song}},
  \bibinfo {author} {\bibfnamefont {M.}~\bibnamefont {Wegner}}, \bibinfo
  {author} {\bibfnamefont {E.}~\bibnamefont {Quandt}}, \bibinfo {author}
  {\bibfnamefont {X.}~\bibnamefont {Chen}}, \emph {et~al.},\ }\bibfield
  {title} {\bibinfo {title} {Power-source-free analysis of pyroelectric energy
  conversion},\ }\href@noop {} {\bibfield  {journal} {\bibinfo  {journal}
  {Phys. Rev. Applied}\ }\textbf {\bibinfo {volume} {12}},\ \bibinfo {pages}
  {014063} (\bibinfo {year} {2019})}\BibitemShut {NoStop}%
\bibitem [{\citenamefont {Ball}\ and\ \citenamefont
  {James}(1987)}]{ball1989fine}%
  \BibitemOpen
  \bibfield  {author} {\bibinfo {author} {\bibfnamefont {J.~M.}\ \bibnamefont
  {Ball}}\ and\ \bibinfo {author} {\bibfnamefont {R.~D.}\ \bibnamefont
  {James}},\ }\bibfield  {title} {\bibinfo {title} {Fine phase mixtures as
  minimizers of energy},\ }\href@noop {} {\bibfield  {journal} {\bibinfo
  {journal} {Arch. Ration. Mech. Anal.}\ }\textbf {\bibinfo {volume} {100}},\
  \bibinfo {pages} {13} (\bibinfo {year} {1987})}\BibitemShut {NoStop}%
\bibitem [{\citenamefont {Chen}\ \emph {et~al.}(2013)\citenamefont {Chen},
  \citenamefont {Srivastava}, \citenamefont {Dabade},\ and\ \citenamefont
  {James}}]{chen2013cc}%
  \BibitemOpen
  \bibfield  {author} {\bibinfo {author} {\bibfnamefont {X.}~\bibnamefont
  {Chen}}, \bibinfo {author} {\bibfnamefont {V.}~\bibnamefont {Srivastava}},
  \bibinfo {author} {\bibfnamefont {V.}~\bibnamefont {Dabade}},\ and\ \bibinfo
  {author} {\bibfnamefont {R.~D.}\ \bibnamefont {James}},\ }\bibfield  {title}
  {\bibinfo {title} {Study of the cofactor conditions: conditions of
  supercompatibility between phases},\ }\href@noop {} {\bibfield  {journal}
  {\bibinfo  {journal} {J. Mech. Phys. Solids}\ }\textbf {\bibinfo {volume}
  {61}},\ \bibinfo {pages} {2566} (\bibinfo {year} {2013})}\BibitemShut
  {NoStop}%
\bibitem [{\citenamefont {Cui}\ \emph {et~al.}(2006)\citenamefont {Cui},
  \citenamefont {Chu}, \citenamefont {Famodu}, \citenamefont {Furuya},
  \citenamefont {Hattrick-Simpers}, \citenamefont {James}, \citenamefont
  {Ludwig}, \citenamefont {Thienhaus}, \citenamefont {Wuttig}, \citenamefont
  {Zhang} \emph {et~al.}}]{cui2006}%
  \BibitemOpen
  \bibfield  {author} {\bibinfo {author} {\bibfnamefont {J.}~\bibnamefont
  {Cui}}, \bibinfo {author} {\bibfnamefont {Y.~S.}\ \bibnamefont {Chu}},
  \bibinfo {author} {\bibfnamefont {O.~O.}\ \bibnamefont {Famodu}}, \bibinfo
  {author} {\bibfnamefont {Y.}~\bibnamefont {Furuya}}, \bibinfo {author}
  {\bibfnamefont {J.}~\bibnamefont {Hattrick-Simpers}}, \bibinfo {author}
  {\bibfnamefont {R.~D.}\ \bibnamefont {James}}, \bibinfo {author}
  {\bibfnamefont {A.}~\bibnamefont {Ludwig}}, \bibinfo {author} {\bibfnamefont
  {S.}~\bibnamefont {Thienhaus}}, \bibinfo {author} {\bibfnamefont
  {M.}~\bibnamefont {Wuttig}}, \bibinfo {author} {\bibfnamefont
  {Z.}~\bibnamefont {Zhang}}, \emph {et~al.},\ }\bibfield  {title} {\bibinfo
  {title} {Combinatorial search of thermoelastic shape-memory alloys with
  extremely small hysteresis width},\ }\href@noop {} {\bibfield  {journal}
  {\bibinfo  {journal} {Nat. Mater.}\ }\textbf {\bibinfo {volume} {5}},\
  \bibinfo {pages} {286} (\bibinfo {year} {2006})}\BibitemShut {NoStop}%
\bibitem [{\citenamefont {Zarnetta}\ \emph {et~al.}(2010)\citenamefont
  {Zarnetta}, \citenamefont {Takahashi}, \citenamefont {Young}, \citenamefont
  {Savan}, \citenamefont {Furuya}, \citenamefont {Thienhaus}, \citenamefont
  {Maa{\ss}}, \citenamefont {Rahim}, \citenamefont {Frenzel}, \citenamefont
  {Brunken} \emph {et~al.}}]{zarnetta2010}%
  \BibitemOpen
  \bibfield  {author} {\bibinfo {author} {\bibfnamefont {R.}~\bibnamefont
  {Zarnetta}}, \bibinfo {author} {\bibfnamefont {R.}~\bibnamefont {Takahashi}},
  \bibinfo {author} {\bibfnamefont {M.~L.}\ \bibnamefont {Young}}, \bibinfo
  {author} {\bibfnamefont {A.}~\bibnamefont {Savan}}, \bibinfo {author}
  {\bibfnamefont {Y.}~\bibnamefont {Furuya}}, \bibinfo {author} {\bibfnamefont
  {S.}~\bibnamefont {Thienhaus}}, \bibinfo {author} {\bibfnamefont
  {B.}~\bibnamefont {Maa{\ss}}}, \bibinfo {author} {\bibfnamefont
  {M.}~\bibnamefont {Rahim}}, \bibinfo {author} {\bibfnamefont
  {J.}~\bibnamefont {Frenzel}}, \bibinfo {author} {\bibfnamefont
  {H.}~\bibnamefont {Brunken}}, \emph {et~al.},\ }\bibfield  {title} {\bibinfo
  {title} {Identification of quaternary shape memory alloys with near-zero
  thermal hysteresis and unprecedented functional stability},\ }\href@noop {}
  {\bibfield  {journal} {\bibinfo  {journal} {Adv. Funct. Mater.}\ }\textbf
  {\bibinfo {volume} {20}},\ \bibinfo {pages} {1917} (\bibinfo {year}
  {2010})}\BibitemShut {NoStop}%
\bibitem [{\citenamefont {Song}\ \emph {et~al.}(2013)\citenamefont {Song},
  \citenamefont {Chen}, \citenamefont {Dabade}, \citenamefont {Shield},\ and\
  \citenamefont {James}}]{song2013}%
  \BibitemOpen
  \bibfield  {author} {\bibinfo {author} {\bibfnamefont {Y.}~\bibnamefont
  {Song}}, \bibinfo {author} {\bibfnamefont {X.}~\bibnamefont {Chen}}, \bibinfo
  {author} {\bibfnamefont {V.}~\bibnamefont {Dabade}}, \bibinfo {author}
  {\bibfnamefont {T.~W.}\ \bibnamefont {Shield}},\ and\ \bibinfo {author}
  {\bibfnamefont {R.~D.}\ \bibnamefont {James}},\ }\bibfield  {title} {\bibinfo
  {title} {Enhanced reversibility and unusual microstructure of a
  phase-transforming material},\ }\href@noop {} {\bibfield  {journal} {\bibinfo
   {journal} {Nature}\ }\textbf {\bibinfo {volume} {502}},\ \bibinfo {pages}
  {85} (\bibinfo {year} {2013})}\BibitemShut {NoStop}%
\bibitem [{\citenamefont {Chluba}\ \emph {et~al.}(2015)\citenamefont {Chluba},
  \citenamefont {Ge}, \citenamefont {de~Miranda}, \citenamefont {Strobel},
  \citenamefont {Kienle}, \citenamefont {Quandt},\ and\ \citenamefont
  {Wuttig}}]{chluba2015}%
  \BibitemOpen
  \bibfield  {author} {\bibinfo {author} {\bibfnamefont {C.}~\bibnamefont
  {Chluba}}, \bibinfo {author} {\bibfnamefont {W.}~\bibnamefont {Ge}}, \bibinfo
  {author} {\bibfnamefont {R.~L.}\ \bibnamefont {de~Miranda}}, \bibinfo
  {author} {\bibfnamefont {J.}~\bibnamefont {Strobel}}, \bibinfo {author}
  {\bibfnamefont {L.}~\bibnamefont {Kienle}}, \bibinfo {author} {\bibfnamefont
  {E.}~\bibnamefont {Quandt}},\ and\ \bibinfo {author} {\bibfnamefont
  {M.}~\bibnamefont {Wuttig}},\ }\bibfield  {title} {\bibinfo {title}
  {Ultralow-fatigue shape memory alloy films},\ }\href@noop {} {\bibfield
  {journal} {\bibinfo  {journal} {Science}\ }\textbf {\bibinfo {volume}
  {348}},\ \bibinfo {pages} {1004} (\bibinfo {year} {2015})}\BibitemShut
  {NoStop}%
\bibitem [{\citenamefont {Pang}\ \emph {et~al.}(2019)\citenamefont {Pang},
  \citenamefont {McCandler},\ and\ \citenamefont {Schuh}}]{pang2019reduced}%
  \BibitemOpen
  \bibfield  {author} {\bibinfo {author} {\bibfnamefont {E.~L.}\ \bibnamefont
  {Pang}}, \bibinfo {author} {\bibfnamefont {C.~A.}\ \bibnamefont
  {McCandler}},\ and\ \bibinfo {author} {\bibfnamefont {C.~A.}\ \bibnamefont
  {Schuh}},\ }\bibfield  {title} {\bibinfo {title} {Reduced cracking in
  polycrystalline {ZrO2-CeO2} shape-memory ceramics by meeting the cofactor
  conditions},\ }\href@noop {} {\bibfield  {journal} {\bibinfo  {journal} {Acta
  Mater.}\ }\textbf {\bibinfo {volume} {177}},\ \bibinfo {pages} {230}
  (\bibinfo {year} {2019})}\BibitemShut {NoStop}%
\bibitem [{\citenamefont {Jetter}\ \emph {et~al.}(2019)\citenamefont {Jetter},
  \citenamefont {Gu}, \citenamefont {Zhang}, \citenamefont {Wuttig},
  \citenamefont {Chen}, \citenamefont {Greer}, \citenamefont {James},\ and\
  \citenamefont {Quandt}}]{jetter2019}%
  \BibitemOpen
  \bibfield  {author} {\bibinfo {author} {\bibfnamefont {J.}~\bibnamefont
  {Jetter}}, \bibinfo {author} {\bibfnamefont {H.}~\bibnamefont {Gu}}, \bibinfo
  {author} {\bibfnamefont {H.}~\bibnamefont {Zhang}}, \bibinfo {author}
  {\bibfnamefont {M.}~\bibnamefont {Wuttig}}, \bibinfo {author} {\bibfnamefont
  {X.}~\bibnamefont {Chen}}, \bibinfo {author} {\bibfnamefont {J.~R.}\
  \bibnamefont {Greer}}, \bibinfo {author} {\bibfnamefont {R.~D.}\ \bibnamefont
  {James}},\ and\ \bibinfo {author} {\bibfnamefont {E.}~\bibnamefont
  {Quandt}},\ }\bibfield  {title} {\bibinfo {title} {Tuning crystallographic
  compatibility to enhance shape memory in ceramics},\ }\href@noop {}
  {\bibfield  {journal} {\bibinfo  {journal} {Phys. Rev. Materials}\ }\textbf
  {\bibinfo {volume} {3}},\ \bibinfo {pages} {093603} (\bibinfo {year}
  {2019})}\BibitemShut {NoStop}%
\bibitem [{\citenamefont {Liang}\ \emph {et~al.}(2020)\citenamefont {Liang},
  \citenamefont {Lee}, \citenamefont {Yu}, \citenamefont {Zhang}, \citenamefont
  {Liang}, \citenamefont {Zavalij}, \citenamefont {Chen}, \citenamefont
  {James}, \citenamefont {Bendersky}, \citenamefont {Davydov} \emph
  {et~al.}}]{liang2020}%
  \BibitemOpen
  \bibfield  {author} {\bibinfo {author} {\bibfnamefont {Y.}~\bibnamefont
  {Liang}}, \bibinfo {author} {\bibfnamefont {S.}~\bibnamefont {Lee}}, \bibinfo
  {author} {\bibfnamefont {H.}~\bibnamefont {Yu}}, \bibinfo {author}
  {\bibfnamefont {H.}~\bibnamefont {Zhang}}, \bibinfo {author} {\bibfnamefont
  {Y.}~\bibnamefont {Liang}}, \bibinfo {author} {\bibfnamefont
  {P.}~\bibnamefont {Zavalij}}, \bibinfo {author} {\bibfnamefont
  {X.}~\bibnamefont {Chen}}, \bibinfo {author} {\bibfnamefont {R.}~\bibnamefont
  {James}}, \bibinfo {author} {\bibfnamefont {L.}~\bibnamefont {Bendersky}},
  \bibinfo {author} {\bibfnamefont {A.}~\bibnamefont {Davydov}}, \emph
  {et~al.},\ }\bibfield  {title} {\bibinfo {title} {Tuning the hysteresis of a
  metal-insulator transition via lattice compatibility},\ }\href@noop {}
  {\bibfield  {journal} {\bibinfo  {journal} {Nat. Comm.}\ }\textbf {\bibinfo
  {volume} {11}},\ \bibinfo {pages} {1} (\bibinfo {year} {2020})}\BibitemShut
  {NoStop}%
\bibitem [{\citenamefont {Wegner}\ \emph {et~al.}(2020)\citenamefont {Wegner},
  \citenamefont {Gu}, \citenamefont {James},\ and\ \citenamefont
  {Quandt}}]{wegner2020}%
  \BibitemOpen
  \bibfield  {author} {\bibinfo {author} {\bibfnamefont {M.}~\bibnamefont
  {Wegner}}, \bibinfo {author} {\bibfnamefont {H.}~\bibnamefont {Gu}}, \bibinfo
  {author} {\bibfnamefont {R.~D.}\ \bibnamefont {James}},\ and\ \bibinfo
  {author} {\bibfnamefont {E.}~\bibnamefont {Quandt}},\ }\bibfield  {title}
  {\bibinfo {title} {Correlation between phase compatibility and efficient
  energy conversion in {Zr-doped Barium Titanate}},\ }\href@noop {} {\bibfield
  {journal} {\bibinfo  {journal} {Sci. Rep.}\ }\textbf {\bibinfo {volume}
  {10}},\ \bibinfo {pages} {1} (\bibinfo {year} {2020})}\BibitemShut {NoStop}%
\bibitem [{\citenamefont {El-Khatib}\ \emph {et~al.}(2019)\citenamefont
  {El-Khatib}, \citenamefont {Bhatti}, \citenamefont {Srivastava},
  \citenamefont {James},\ and\ \citenamefont {Leighton}}]{el2019PRM}%
  \BibitemOpen
  \bibfield  {author} {\bibinfo {author} {\bibfnamefont {S.}~\bibnamefont
  {El-Khatib}}, \bibinfo {author} {\bibfnamefont {K.~P.}\ \bibnamefont
  {Bhatti}}, \bibinfo {author} {\bibfnamefont {V.}~\bibnamefont {Srivastava}},
  \bibinfo {author} {\bibfnamefont {R.}~\bibnamefont {James}},\ and\ \bibinfo
  {author} {\bibfnamefont {C.}~\bibnamefont {Leighton}},\ }\bibfield  {title}
  {\bibinfo {title} {Nanoscale magnetic phase competition throughout the
  {Ni$_{50-x}$ Co$_x$Mn$_{40}$Sn$_{10}$} phase diagram: Insights from
  small-angle neutron scattering},\ }\href@noop {} {\bibfield  {journal}
  {\bibinfo  {journal} {Phys. Rev. Materials}\ }\textbf {\bibinfo {volume}
  {3}},\ \bibinfo {pages} {104413} (\bibinfo {year} {2019})}\BibitemShut
  {NoStop}%
\bibitem [{\citenamefont {Zhao}\ \emph {et~al.}(2017)\citenamefont {Zhao},
  \citenamefont {Liu}, \citenamefont {Chen}, \citenamefont {Sun}, \citenamefont
  {Li}, \citenamefont {Zhang}, \citenamefont {Shao}, \citenamefont {Zhang},\
  and\ \citenamefont {Yan}}]{zhao2017acta}%
  \BibitemOpen
  \bibfield  {author} {\bibinfo {author} {\bibfnamefont {D.}~\bibnamefont
  {Zhao}}, \bibinfo {author} {\bibfnamefont {J.}~\bibnamefont {Liu}}, \bibinfo
  {author} {\bibfnamefont {X.}~\bibnamefont {Chen}}, \bibinfo {author}
  {\bibfnamefont {W.}~\bibnamefont {Sun}}, \bibinfo {author} {\bibfnamefont
  {Y.}~\bibnamefont {Li}}, \bibinfo {author} {\bibfnamefont {M.}~\bibnamefont
  {Zhang}}, \bibinfo {author} {\bibfnamefont {Y.}~\bibnamefont {Shao}},
  \bibinfo {author} {\bibfnamefont {H.}~\bibnamefont {Zhang}},\ and\ \bibinfo
  {author} {\bibfnamefont {A.}~\bibnamefont {Yan}},\ }\bibfield  {title}
  {\bibinfo {title} {Giant caloric effect of low-hysteresis metamagnetic shape
  memory alloys with exceptional cyclic functionality},\ }\href@noop {}
  {\bibfield  {journal} {\bibinfo  {journal} {Acta Mater.}\ }\textbf {\bibinfo
  {volume} {133}},\ \bibinfo {pages} {217} (\bibinfo {year}
  {2017})}\BibitemShut {NoStop}%
\bibitem [{\citenamefont {Liu}\ \emph {et~al.}(2015)\citenamefont {Liu},
  \citenamefont {Wu}, \citenamefont {Chen}, \citenamefont {Fang}, \citenamefont
  {Ding}, \citenamefont {Zhao},\ and\ \citenamefont {Luo}}]{liu2015BCZT}%
  \BibitemOpen
  \bibfield  {author} {\bibinfo {author} {\bibfnamefont {X.}~\bibnamefont
  {Liu}}, \bibinfo {author} {\bibfnamefont {D.}~\bibnamefont {Wu}}, \bibinfo
  {author} {\bibfnamefont {Z.}~\bibnamefont {Chen}}, \bibinfo {author}
  {\bibfnamefont {B.}~\bibnamefont {Fang}}, \bibinfo {author} {\bibfnamefont
  {J.}~\bibnamefont {Ding}}, \bibinfo {author} {\bibfnamefont {X.}~\bibnamefont
  {Zhao}},\ and\ \bibinfo {author} {\bibfnamefont {H.}~\bibnamefont {Luo}},\
  }\bibfield  {title} {\bibinfo {title} {Ferroelectric, dielectric and
  pyroelectric properties of {Sr} and {Sn} codoped {BCZT} lead free ceramics},\
  }\href@noop {} {\bibfield  {journal} {\bibinfo  {journal} {Advances in
  Applied Ceramics}\ }\textbf {\bibinfo {volume} {114}},\ \bibinfo {pages}
  {436} (\bibinfo {year} {2015})}\BibitemShut {NoStop}%
\bibitem [{\citenamefont {Coondoo}\ \emph {et~al.}(2021)\citenamefont
  {Coondoo}, \citenamefont {Panwar}, \citenamefont {Krylova}, \citenamefont
  {Krylov}, \citenamefont {Alikin}, \citenamefont {Jakka}, \citenamefont
  {Turygin}, \citenamefont {Shur},\ and\ \citenamefont
  {Kholkin}}]{coondoo2021}%
  \BibitemOpen
  \bibfield  {author} {\bibinfo {author} {\bibfnamefont {I.}~\bibnamefont
  {Coondoo}}, \bibinfo {author} {\bibfnamefont {N.}~\bibnamefont {Panwar}},
  \bibinfo {author} {\bibfnamefont {S.}~\bibnamefont {Krylova}}, \bibinfo
  {author} {\bibfnamefont {A.}~\bibnamefont {Krylov}}, \bibinfo {author}
  {\bibfnamefont {D.}~\bibnamefont {Alikin}}, \bibinfo {author} {\bibfnamefont
  {S.~K.}\ \bibnamefont {Jakka}}, \bibinfo {author} {\bibfnamefont
  {A.}~\bibnamefont {Turygin}}, \bibinfo {author} {\bibfnamefont {V.~Y.}\
  \bibnamefont {Shur}},\ and\ \bibinfo {author} {\bibfnamefont {A.~L.}\
  \bibnamefont {Kholkin}},\ }\bibfield  {title} {\bibinfo {title}
  {Temperature-dependent {Raman} spectroscopy, domain morphology and
  photoluminescence studies in lead-free {BCZT} ceramic},\ }\href@noop {}
  {\bibfield  {journal} {\bibinfo  {journal} {Ceramics International}\ }\textbf
  {\bibinfo {volume} {47}},\ \bibinfo {pages} {2828} (\bibinfo {year}
  {2021})}\BibitemShut {NoStop}%
\bibitem [{\citenamefont {Zhang}\ \emph
  {et~al.}(2020{\natexlab{b}})\citenamefont {Zhang}, \citenamefont {Zeng},
  \citenamefont {Zhu}, \citenamefont {Karami}, \citenamefont {Chen} \emph
  {et~al.}}]{zhang2020leakage}%
  \BibitemOpen
  \bibfield  {author} {\bibinfo {author} {\bibfnamefont {C.}~\bibnamefont
  {Zhang}}, \bibinfo {author} {\bibfnamefont {Z.}~\bibnamefont {Zeng}},
  \bibinfo {author} {\bibfnamefont {Z.}~\bibnamefont {Zhu}}, \bibinfo {author}
  {\bibfnamefont {M.}~\bibnamefont {Karami}}, \bibinfo {author} {\bibfnamefont
  {X.}~\bibnamefont {Chen}}, \emph {et~al.},\ }\bibfield  {title} {\bibinfo
  {title} {Impact of leakage for electricity generation by pyroelectric
  converter},\ }\href@noop {} {\bibfield  {journal} {\bibinfo  {journal} {Phys.
  Rev. Applied}\ }\textbf {\bibinfo {volume} {14}},\ \bibinfo {pages} {064079}
  (\bibinfo {year} {2020}{\natexlab{b}})}\BibitemShut {NoStop}%
\bibitem [{\citenamefont {Nagaraj}\ \emph {et~al.}(1999)\citenamefont
  {Nagaraj}, \citenamefont {Aggarwal}, \citenamefont {Song}, \citenamefont
  {Sawhney},\ and\ \citenamefont {Ramesh}}]{nagaraj1999leakage}%
  \BibitemOpen
  \bibfield  {author} {\bibinfo {author} {\bibfnamefont {B.}~\bibnamefont
  {Nagaraj}}, \bibinfo {author} {\bibfnamefont {S.}~\bibnamefont {Aggarwal}},
  \bibinfo {author} {\bibfnamefont {T.}~\bibnamefont {Song}}, \bibinfo {author}
  {\bibfnamefont {T.}~\bibnamefont {Sawhney}},\ and\ \bibinfo {author}
  {\bibfnamefont {R.}~\bibnamefont {Ramesh}},\ }\bibfield  {title} {\bibinfo
  {title} {Leakage current mechanisms in lead-based thin-film ferroelectric
  capacitors},\ }\href@noop {} {\bibfield  {journal} {\bibinfo  {journal}
  {Physical Review B}\ }\textbf {\bibinfo {volume} {59}},\ \bibinfo {pages}
  {16022} (\bibinfo {year} {1999})}\BibitemShut {NoStop}%
\bibitem [{\citenamefont {Pabst}\ \emph {et~al.}(2007)\citenamefont {Pabst},
  \citenamefont {Martin}, \citenamefont {Chu},\ and\ \citenamefont
  {Ramesh}}]{pabst2007leakage}%
  \BibitemOpen
  \bibfield  {author} {\bibinfo {author} {\bibfnamefont {G.~W.}\ \bibnamefont
  {Pabst}}, \bibinfo {author} {\bibfnamefont {L.~W.}\ \bibnamefont {Martin}},
  \bibinfo {author} {\bibfnamefont {Y.-H.}\ \bibnamefont {Chu}},\ and\ \bibinfo
  {author} {\bibfnamefont {R.}~\bibnamefont {Ramesh}},\ }\bibfield  {title}
  {\bibinfo {title} {Leakage mechanisms in {BiFeO3} thin films},\ }\href@noop
  {} {\bibfield  {journal} {\bibinfo  {journal} {Applied Physics Letters}\
  }\textbf {\bibinfo {volume} {90}},\ \bibinfo {pages} {072902} (\bibinfo
  {year} {2007})}\BibitemShut {NoStop}%
\bibitem [{\citenamefont {Wang}\ \emph {et~al.}(2005)\citenamefont {Wang},
  \citenamefont {Cheng}, \citenamefont {Wang}, \citenamefont {Redfern},
  \citenamefont {Dai}, \citenamefont {Jin}, \citenamefont {Lu}, \citenamefont
  {Zhou}, \citenamefont {Chen},\ and\ \citenamefont {Yang}}]{wang2005leak}%
  \BibitemOpen
  \bibfield  {author} {\bibinfo {author} {\bibfnamefont {S.}~\bibnamefont
  {Wang}}, \bibinfo {author} {\bibfnamefont {B.}~\bibnamefont {Cheng}},
  \bibinfo {author} {\bibfnamefont {C.}~\bibnamefont {Wang}}, \bibinfo {author}
  {\bibfnamefont {S.}~\bibnamefont {Redfern}}, \bibinfo {author} {\bibfnamefont
  {S.}~\bibnamefont {Dai}}, \bibinfo {author} {\bibfnamefont {K.}~\bibnamefont
  {Jin}}, \bibinfo {author} {\bibfnamefont {H.}~\bibnamefont {Lu}}, \bibinfo
  {author} {\bibfnamefont {Y.}~\bibnamefont {Zhou}}, \bibinfo {author}
  {\bibfnamefont {Z.}~\bibnamefont {Chen}},\ and\ \bibinfo {author}
  {\bibfnamefont {G.}~\bibnamefont {Yang}},\ }\bibfield  {title} {\bibinfo
  {title} {Influence of {Ce} doping on leakage current in
  {Ba$_{0.5}$Sr$_{0.5}$TiO$_3$} films},\ }\href@noop {} {\bibfield  {journal}
  {\bibinfo  {journal} {J. Phys. D: Appl. Phys.}\ }\textbf {\bibinfo {volume}
  {38}},\ \bibinfo {pages} {2253} (\bibinfo {year} {2005})}\BibitemShut
  {NoStop}%
\bibitem [{\citenamefont {Balbashov}\ and\ \citenamefont
  {Egorov}(1981)}]{balbashov1981}%
  \BibitemOpen
  \bibfield  {author} {\bibinfo {author} {\bibfnamefont {A.~M.}\ \bibnamefont
  {Balbashov}}\ and\ \bibinfo {author} {\bibfnamefont {S.~K.}\ \bibnamefont
  {Egorov}},\ }\bibfield  {title} {\bibinfo {title} {Apparatus for growth of
  single crystals of oxide compounds by floating zone melting with radiation
  heating},\ }\href@noop {} {\bibfield  {journal} {\bibinfo  {journal} {Journal
  of Crystal Growth}\ }\textbf {\bibinfo {volume} {52}},\ \bibinfo {pages}
  {498} (\bibinfo {year} {1981})}\BibitemShut {NoStop}%
\bibitem [{\citenamefont {Kimura}\ and\ \citenamefont
  {Kitamura}(1992)}]{kimura1992}%
  \BibitemOpen
  \bibfield  {author} {\bibinfo {author} {\bibfnamefont {S.}~\bibnamefont
  {Kimura}}\ and\ \bibinfo {author} {\bibfnamefont {K.}~\bibnamefont
  {Kitamura}},\ }\bibfield  {title} {\bibinfo {title} {{Floating zone crystal
  growth and phase equilibria: A review}},\ }\href@noop {} {\bibfield
  {journal} {\bibinfo  {journal} {Journal of the American Ceramic Society}\
  }\textbf {\bibinfo {volume} {75}},\ \bibinfo {pages} {1440} (\bibinfo {year}
  {1992})}\BibitemShut {NoStop}%
\bibitem [{\citenamefont {Lau}\ \emph {et~al.}(2008)\citenamefont {Lau},
  \citenamefont {Cheng}, \citenamefont {Choy}, \citenamefont {Lin},
  \citenamefont {Kwok},\ and\ \citenamefont {Chan}}]{lau2008lead}%
  \BibitemOpen
  \bibfield  {author} {\bibinfo {author} {\bibfnamefont {S.~T.}\ \bibnamefont
  {Lau}}, \bibinfo {author} {\bibfnamefont {C.}~\bibnamefont {Cheng}}, \bibinfo
  {author} {\bibfnamefont {S.}~\bibnamefont {Choy}}, \bibinfo {author}
  {\bibfnamefont {D.}~\bibnamefont {Lin}}, \bibinfo {author} {\bibfnamefont
  {K.~W.}\ \bibnamefont {Kwok}},\ and\ \bibinfo {author} {\bibfnamefont
  {H.~L.}\ \bibnamefont {Chan}},\ }\bibfield  {title} {\bibinfo {title}
  {Lead-free ceramics for pyroelectric applications},\ }\href@noop {}
  {\bibfield  {journal} {\bibinfo  {journal} {Journal of applied physics}\
  }\textbf {\bibinfo {volume} {103}},\ \bibinfo {pages} {104105} (\bibinfo
  {year} {2008})}\BibitemShut {NoStop}%
\bibitem [{\citenamefont {Chen}\ \emph {et~al.}(2018)\citenamefont {Chen},
  \citenamefont {Kang}, \citenamefont {Lu}, \citenamefont {Luo}, \citenamefont
  {Wang},\ and\ \citenamefont {He}}]{chen2018general}%
  \BibitemOpen
  \bibfield  {author} {\bibinfo {author} {\bibfnamefont {J.}~\bibnamefont
  {Chen}}, \bibinfo {author} {\bibfnamefont {L.}~\bibnamefont {Kang}}, \bibinfo
  {author} {\bibfnamefont {H.}~\bibnamefont {Lu}}, \bibinfo {author}
  {\bibfnamefont {P.}~\bibnamefont {Luo}}, \bibinfo {author} {\bibfnamefont
  {F.}~\bibnamefont {Wang}},\ and\ \bibinfo {author} {\bibfnamefont
  {L.}~\bibnamefont {He}},\ }\bibfield  {title} {\bibinfo {title} {The general
  purpose powder diffractometer at {CSNS}},\ }\href@noop {} {\bibfield
  {journal} {\bibinfo  {journal} {Physica B: Condensed Matter}\ }\textbf
  {\bibinfo {volume} {551}},\ \bibinfo {pages} {370} (\bibinfo {year}
  {2018})}\BibitemShut {NoStop}%
\bibitem [{\citenamefont {Toby}(2001)}]{toby2001expgui}%
  \BibitemOpen
  \bibfield  {author} {\bibinfo {author} {\bibfnamefont {B.~H.}\ \bibnamefont
  {Toby}},\ }\bibfield  {title} {\bibinfo {title} {{EXPGUI, a graphical user
  interface for GSAS}},\ }\href@noop {} {\bibfield  {journal} {\bibinfo
  {journal} {Journal of Applied Crystallography}\ }\textbf {\bibinfo {volume}
  {34}},\ \bibinfo {pages} {210} (\bibinfo {year} {2001})}\BibitemShut
  {NoStop}%
\bibitem [{\citenamefont {Zhang}\ \emph {et~al.}(2009)\citenamefont {Zhang},
  \citenamefont {James},\ and\ \citenamefont {M{\"u}ller}}]{zhang2009energy}%
  \BibitemOpen
  \bibfield  {author} {\bibinfo {author} {\bibfnamefont {Z.}~\bibnamefont
  {Zhang}}, \bibinfo {author} {\bibfnamefont {R.~D.}\ \bibnamefont {James}},\
  and\ \bibinfo {author} {\bibfnamefont {S.}~\bibnamefont {M{\"u}ller}},\
  }\bibfield  {title} {\bibinfo {title} {Energy barriers and hysteresis in
  martensitic phase transformations},\ }\href@noop {} {\bibfield  {journal}
  {\bibinfo  {journal} {Acta Materialia}\ }\textbf {\bibinfo {volume} {57}},\
  \bibinfo {pages} {4332} (\bibinfo {year} {2009})}\BibitemShut {NoStop}%
\end{thebibliography}%

\end{document}